\documentclass[10pt
]{article}
\usepackage[
textsize=tiny, backgroundcolor=white, linecolor=red, bordercolor=red]{todonotes}

\usepackage[bookmarks, colorlinks=TRUE, linkcolor=blue, urlcolor=blue,
citecolor=black, pdftitle={Soft Maximin Estimation for Heterogeneous Data}, pdfauthor={Adam Lund}]{hyperref}
\usepackage[textsize=tiny, backgroundcolor=white, linecolor=red, bordercolor=red]{todonotes}
\usepackage{lineno}
\usepackage{setspace}
\usepackage{mathpazo,mathrsfs}
\usepackage[sort&compress, authoryear]{natbib}
\usepackage{mathtools}
\usepackage{enumerate}

 \usepackage{rotating}
\usepackage{graphicx}

\usepackage[margin = 1.875in]{geometry} 

\usepackage{amsmath}
\usepackage{amssymb}
\usepackage{amsthm}
\usepackage{algorithm}
\usepackage{algorithmic}
\usepackage{float}
 \usepackage{mathabx}

\newcommand{\NN}{\mathbb{N}}

\newcommand{\RR}{\mathbb{R}}

\newcommand{\lse}{\mathrm{lse}}
\newcommand{\supp}{\mathop{\mathrm{supp}}}

\newcommand{\bs}[1]{\boldsymbol{#1}}
\DeclareMathOperator*{\argmax}{arg\,max}
\DeclareMathOperator*{\argmin}{arg\,min}

\newtheorem{thm_lemma}{Lemma}
\newtheorem{thm_cor}{Corollary}

\newtheorem{thm_prop}{Proposition}

\begin{document}

\title{Soft Maximin Estimation for   Heterogeneous  Data}
\author{Adam Lund\thanks{Speevr Labs, Speevr, London, e-mail: adam.lund@math.ku.dk or @speevr.com.},  
 S\o{}ren Wengel Mogensen\thanks{University of Copenhagen, Department of Mathematical Sciences}, 
 and Niels Richard Hansen\thanks{University of Copenhagen, Department of Mathematical Sciences}}

\maketitle
\begin{abstract}
Extracting a common robust signal from data 
 divided into heterogeneous groups can be difficult when each group -- in addition to the signal -- can contain large, unique variation components. Previously, maximin estimation has been proposed as a robust estimation method in the presence of  heterogeneous noise. We propose soft maximin estimation as
a computationally attractive  alternative  aimed at striking a balance between pooled estimation and (hard) maximin estimation. 

The soft maximin method provides a range of estimators, controlled by a parameter $\zeta>0$, that interpolates pooled  least squares estimation and maximin estimation. By establishing relevant theoretical properties we argue that   the soft maximin method is  both statistically sensibel and computationally attractive.   

We  also demonstrate, on    real  and simulated  data, that the soft maximin estimator  can offer improvements over both pooled OLS and hard maximin  in terms of predictive performance and computational complexity.

A time and memory efficient implementation is provided in the  R package \verb+SMME+ available on CRAN.
\end{abstract}

  heterogeneous data, robust estimation, regularization, convex optimization

\section{Introduction}\label{sec:one}
 We consider the problem of extracting a common signal  from heterogeneous  data. As  heterogeneity is prevalent in large-scale settings  our  aim is  a computationally efficient estimator (solution) with good
 statistical properties under varying degrees of  data heterogeneity.

To make the concept of heterogeneity concrete, consider the linear   \emph{mixture} model with univariate response variables $Y_1,\ldots,Y_n$   generated as
\begin{alignat}{4}\label{eq1}
  Y_{ i} =X_{ i}^\top B_i + \varepsilon_{i}, \quad i =  1, \ldots, n.
\end{alignat}
Here, $B_1,\dots, B_n$ and the feature vectors $X_{1},\ldots,X_{n}$ are $p$-dimensional random variables
and $\varepsilon_{1}, \ldots, \varepsilon_{n}$ are univariate noise variables.
The feature vectors are observed and assumed i.i.d., and
the noise variables are likewise i.i.d. The unobserved variables
$B_1,\dots, B_n$ are identically distributed with distribution $F_B$ but
not necessarily independent,  see also \cite{meinshausen2015}.

Heterogeneity in the model given by \eqref{eq1} is due to the variation in $B_i$
as governed by $F_B$. Because the $B_i$-s can be dependent, model \eqref{eq1}
can capture heterogeneity caused by a group structure, that is, when data comes
with a natural grouping and $B_i$ is constant within groups but vary across
groups. Even if data is not grouped, or if the group structure is unknown, it is
beneficial to study the setup with a known group structure. In the example below
on bike sharing, a group structure is introduced to represent temporal
heterogeneity, and \cite{meinshausen2015} demonstrate how to construct group
structures as part of the inference when no grouping is given.

 Our focus is therefore on a setup with $G$ groups and with $B_i$ constant within groups. The objective is to learn a single $\beta \in \mathbb{R}^p$ that
 can sensibly be regarded as a \emph{common signal} of the $B_i$-s. Pooling
 data across groups and computing the OLS estimator may be non-robust,
 depending on $F_B$, and \cite{meinshausen2015} introduced maximin estimation
 as a robust alternative to OLS for heterogeneous data from the model \eqref{eq1}.
 The common signal estimated by maximin estimation is the population quantity
 called the maximin effect.
 
\begin{figure}[H]
\centering
\includegraphics[scale=0.4]{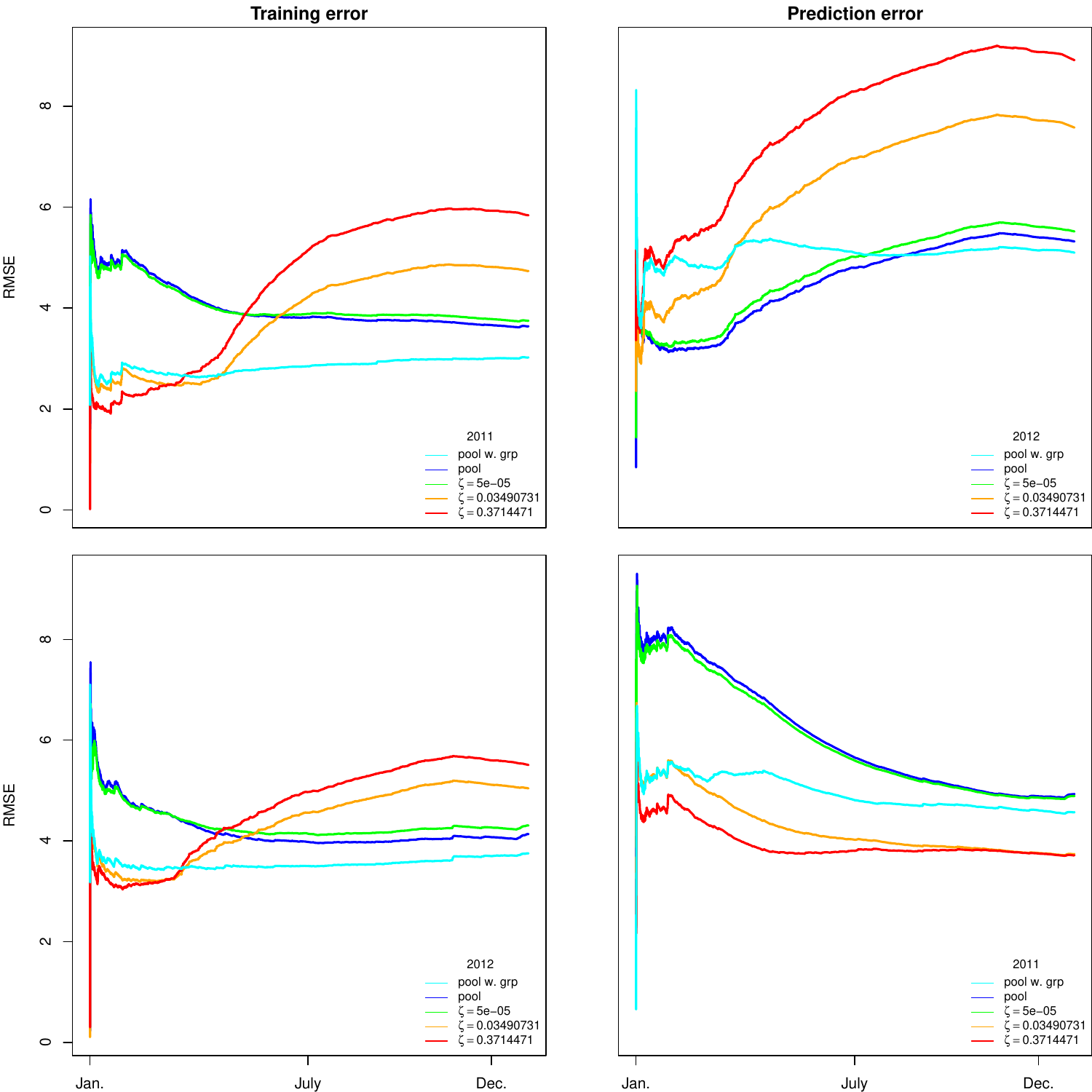}
\caption{Top: RMSE computed for 2011 training data and 2012 validation data for the soft maximin estimators for 3 values of $\zeta$  (resp. green, orange and red), the pooled OLS estimator (blue) and the pooled OLS estimator using the group variable \texttt{mnth} (cyan). Bottom: RMSE computed for 2012 training data and 2011 validation data using the same methods and coloring.}
\label{fig:1new}
\end{figure}

While the maximin estimator is robust, it can also be conservative, and we
propose \emph{soft maximin estimation} to strike a good balance between
maximin and pooled OLS estimation. The balance is controlled by a tuning
parameter $\zeta>0$, with $\zeta \to \infty$ corresponding to maximin estimation.
Figure \ref{fig:1new} shows the result of applying  the soft maximin estimator, for three values of $\zeta$,  as well as  the pooled OLS estimator  to a real data set. It illustrates how predictive performance of the two
extreme estimators is interpolated by soft maximin estimation, quantified as cumulative root mean square error over time, see \eqref{eq21new}.

The specific application illustrated in Figure \ref{fig:1new} is described in detail in Section \ref{sec:bike}. The data is  on the hourly number of bike shares (see \cite{fanaee-t2013}) from two years (2011 and 2012), and the model predicts this number based on weekday, time of the day and the weather. In this application,
one year is used for training and the other year is used for validation (prediction). To safeguard against  temporal heterogeneity the data is grouped according to  the months variable \verb+mnth+, thus $G = 12$. Training on 2011 makes soft maximin with a high value of $\zeta$ too conservative, which leads to poor predictive performance. In this
case, pooled OLS or soft maximin with a low $\zeta$ perform best. However,
training on 2012 data makes the pooled OLS estimator overfit and soft maximin
with a large $\zeta$ has better predictive performance.
The paper is organized as follows. The   model and estimation framework is outlined in Section
 \ref{sec:softmaximin} and   statistical properties of   soft maximin estimation  are discussed. This section includes theoretical results supporting that
 soft maximin interpolates maximin and pooled OLS estimation.
 In Section  \ref{sec:three} we propose an algorithm for computing the general soft maximin estimator. The algorithm solves a non-differentiable convex optimization problem, but as opposed to maximin estimation \citep{meinshausen2015}, the
 problem we solve is separable in the sense of \cite{tseng2009}. This makes
 it notably easier to construct efficient algorithms with convergence guarantees.
 Section \ref{sec:three} also includes theoretical bounds on Lipschitz constants,
 which can be used to select an efficient  step size in our solution algorithm,
 and a discussion of how our results can be applied efficiently to array tensor smoothing.
 In Section \ref{sec:four}, we present the application to the bike share data
 and results from a simulation study on array tensor smoothing. The simulation
 study was inspired by the application to neuronal activity data analyzed in Appendix \ref{sec:neuron}. Section \ref{sec:five}  summarizes    the proposed methodology  and its relation to alternative methods. The algorithms are implemented in the R package \verb+SMME+ available from CRAN, see \cite{lund2021cran}.

\section{Soft maximin estimation}\label{sec:softmaximin}
Here   we  present the methodology    in a setup with a  given group structure and  effects $B_i$  constant within  each group. This in turn implies a finite support for $F_B$ and this model is  perhaps better understood as a type of linear \emph{mixed} model as the  grouping is available  hence no longer part of the inference. In contrast to a traditional mixed model, however, we avoid explicitly modeling fixed and random effects since the aim is not to draw inference about these. Instead we seek to obtain only  an estimate of the possible common effects present in the data.

To introduce $G\in \NN$ groups for the model \eqref{eq1},    suppose given a partition  $I_1, \ldots ,I_G$  of the index set $\{1,\ldots, n\}$ such that $\vert I_g\vert=n_g$, and  $n=\sum_gn_g$ where for each $g$ and for all $i\in I_g$, $B_i=B_g$ is  the  effect in group $g$. Thus $F_B$ has finite unknown support with cardinality $G$, that is $\supp(F_B)=\{b_1,\ldots,b_G \}\subset \RR^p$ with $b_g:=B_g(\omega)$  the true unknown effect in group  $g$.

Using this extra structure we can label the   response, covariates and errors according to  group. For  the $g$th group let $\bs{Y}_g = (Y_{g, 1},\ldots,Y_{g, n_g})^\top$  be  the $n_g\times 1$ response vector, $\bs{X}_g =(X_{g,  1}\vert \ldots \vert X_{g, n_g})^\top$  the  $n_g\times p$  design matrix, and $\bs \varepsilon_g =
(\varepsilon_{g,1},\ldots,\varepsilon_{g,n_g})^\top$  the  $n_g\times 1$ error vector. The
linear model for the $g$th group is then
 \begin{alignat}{4}\label{eq5}
  \bs Y_g=\bs X_gb_g+\bs \varepsilon_g,\quad g\in \{1,\ldots,G\}.
\end{alignat}
A \emph{common signal} in this framework is represented by a single
$\beta \in \mathbb{R}^p$ such that $\bs X_g \beta$ is a good and robust approximation
of $\bs X_gb_g$ across all $G$ groups.

To   gauge the quality of the  approximation we adopt the optimality criterion from   \cite{meinshausen2015}. There the explained variance in group $g$ when using some $\beta\in \RR^p$ in \eqref{eq5} is  defined as
\begin{alignat}{4}\label{eq3}
  V_{b_g}(\beta):=  2\beta^\top\Sigma b_g-\beta^\top\Sigma \beta.
\end{alignat}
 The optimal approximation across all groups is then the so-called maximin  effect  defined as a $b^\ast\in \RR^p$ that maximizes the minimum of the explained variances across groups, i.e.  $b^\ast:=\argmax_\beta \min_gV_{b_g}(\beta)$.

Since $b_g$ is unknown, to make this criteria operational, let $\hat \Sigma_g:=\bs X_g^\top\bs X_g/n_g$ denote the empirical Gram matrix in group $g$. By \eqref{eq5}  replacing $\Sigma$ with $\hat \Sigma_g $ in \eqref{eq3} we obtain  the empirical explained variance in group $g$
\begin{alignat}{4}\label{eq9}
    \hat V_g(\beta)  \coloneqq \frac{1}{n_g}(2\beta^\top \bs{X}_g^\top \bs Y_g-\beta^\top \bs{X}_g^\top\bs{X}_g\beta).
 \end{alignat}
 The maximin effects estimator  is  obtained by  maximizing (the  penalized) minimum of \eqref{eq9} or equivalently by minimizing (the penalized)  maximin loss function   $\beta\mapsto \max_g\{-\hat V_{g}(\beta)\}$, see \eqref{eq:maximin} in Section \ref{sec:five}.

 As shown    in \cite{meinshausen2015} the use of this estimator  can lead to more robust estimates for heterogeneous data compared to an estimator that does not take grouping into account i.e. a   pooled estimator. The intuition is that the maximin estimator extracts only features that are active with the same sign across groups while setting group specific features to zero. This makes it a more crude  estimator compared to one obtained using the full mixed model methodology, however it is in principle  also more robust and  potentially computationally more attractive. 
  In large scale data settings, where  data heterogeneity is typically encountered, the computational aspect of the estimator is crucial. However, since the $\max$-function is non-differentiable and non-separable     the maximin problem \eqref{eq:maximin} is not easy to solve.

We  address this computational hurdle  by replacing the  $\max$-function with the following smooth function. For $G\in \NN$ and  $\zeta \neq 0$ consider the scaled log-sum exponential function
\begin{alignat}{4}\label{eq5new}
\lse_\zeta(x) \coloneqq \frac {\log(\sum_j  e^{\zeta x_j} )}{\zeta}, \quad x\in
\RR^G.
\end{alignat}
Clearly $\lse_\zeta$ is differentiable and as we  show in Section \ref{sec:three} it has additional  properties that makes it  well suited for optimization purposes. First,  the  basic properties stated next are easily verified (see the appendix) and highlight why \eqref{eq5new} is a sensible choice  as an approximation of the   $\max$-function.\begin{thm_lemma}\label{lemma:0}
Let $G\in \NN$ and $x\in\RR^G$.
\begin{enumerate}[i)]
\item For $\zeta >0 $  it holds that
\begin{alignat}{4}
  \max\{x_1,\ldots,x_G\}\leq \lse_\zeta(x)\leq  \frac {\log(G  )}{\zeta}+ \max\{x_1,\ldots,x_G\},
\end{alignat}
and in particular  $\lse_{\zeta}(x)\searrow  \max_g \{x_g\}  $ as $\zeta \to \infty$.
\item For  $\zeta\to 0$   it holds that
$$\lse_{\zeta}(x)=  \frac{1}{G}\sum_{j=1}^Gx_j + \frac{\log(G)}{\zeta} +o(1).$$
\end{enumerate}
\end{thm_lemma}

We define the soft  maximin loss function, by
\begin{alignat*}{4}
s_{\zeta}(\beta) \coloneqq \lse_{\zeta}(-\hat V(\beta)),\quad \beta\in \RR^p,\quad \zeta > 0,
\end{alignat*}
where $\hat V(\beta) := (\hat V_1(\beta),\ldots, \hat V_G(\beta))^\top$.  For   $\kappa>0$ and $\zeta>0$,
 the soft maximin estimator   can now be defined by
\begin{alignat}{4}\label{eq21}
\hat \beta^\kappa_{smm} :=\arg \min_{\beta}   \lse_\zeta(-\hat V(\beta)) \quad \text{s.t.}\quad \Vert\beta\Vert_1\leq\kappa.
\end{alignat}

Using Lemma \ref{lemma:0}, it is possible   to quantify  the impact of  the parameter $\zeta$ on the performance of the soft maximin estimator \eqref{eq21}. The following result   gives a bound on the  maximum  negative explained variance of the soft maximin estimator,  using that of the   theoretical maximin effect $b^\ast$.
\begin{thm_prop}\label{prop:zero}
 Let $D = \max_g \Vert \hat \Sigma_g - \Sigma \Vert_\infty$   and
 $\delta=\max_g\Vert \bs X_g^\top \bs\varepsilon_g/n_g\Vert_\infty$. For fixed $\zeta>0$ and $ \kappa>0$, if $\kappa  \geq \max_g \Vert b_g\Vert_1$,
 \begin{alignat*}{4}
\max_g\{ -V_{b_g}( \hat\beta_{smm}^\kappa )\}
 \leq  \max_g\{-V_{b_g}(b^\ast)\} +6D\kappa ^2 +4  \kappa\delta +\frac{\log(G)}{\zeta},
\end{alignat*}
where $b^\ast$ is the maximin effect. In particular
 \begin{alignat*}{4}
\Vert \hat \beta_{smm}^\kappa - b^\ast\Vert_\Sigma
 \leq6\kappa^2 D +4\kappa \delta+\frac{\log(G)}{\zeta}.
\end{alignat*}
\end{thm_prop}

 Proposition \ref{prop:zero} is  shown  by combining  Lemma \ref{lemma:0} and
    results  in \cite{meinshausen2015}, see the appendix. In particular, the performance loss incurred when using  the soft maximin estimator is  bounded by the same quantity as  that of the maximin estimator  plus  the soft maximum approximation bias   $\log(G)/\zeta$ from Lemma \ref{lemma:0}.  Thus the soft maximin estimator enjoys  theoretical properties similar to those of the (hard) maximin estimator, when controlling for the parameter $\zeta$. In particular,  for $D=0$ (e.g. for a fixed design) and a fixed number of groups, if  $n_g\to \infty$ for all $g$,  the soft maximin estimator only retains the approximation bias.

 Proposition \ref{prop:zero}   establishes a connection between the soft maximin performance and the maximin effect and shows that  for $\zeta\uparrow\infty$ we indeed  obtain the maximin estimator performance. However,  it also highlights that for  $\zeta  \downarrow0$ the performance of the soft maximin estimator can stray arbitrarily far away   from that of the maximin estimator.  To shed light on this note that by Lemma \ref{lemma:0}, for small $\zeta>0$,
\begin{alignat*}{4}
  s_{\zeta}(\beta)\approx\frac{1}{G}\sum_{j=1}^G-\hat V_j(\beta) +\frac{\log(G)}{\zeta}\propto\frac{1}{n}\sum_{j=1}^G\sum_{i=1}^{n_j}\frac{n}{Gn_j}( (X_{j}\beta)_i- Y_{j,i})^2
 \end{alignat*}
 and \eqref{eq21} effectively becomes  a penalized weighted least squares (PWLS) problem over all $n$ observations. Thus solving \eqref{eq21} for a small $\zeta  \geq0$  approximately yields the  pooled PWLS estimator with weights  amplifying  observations from smaller  than average  groups. With the same number of observations in each group,   the soft maximin estimator in turn   interpolates the  pooled PLS estimator and the maximin estimator.

 In  this sense $\zeta $ reflects the heterogeneity in the data. If there is little heterogeneity a low or even zero $\zeta$ might work well corresponding to  grouping not being relevant.  However, for  heterogeneous data a low $\zeta$ might lead to predictions that are worse than the zero prediction, whereas a high $\zeta$ can still work well.
 
 To illustrate the interpolation,  consider a small data  example with data  generated according to \eqref{eq5}  with $G=20$ groups,  $n_g = 400$ observations in each group, and a two-dimensional parameter  space. For fixed effects   $\{b_1,\ldots, b_{20}\}\subset\ \RR^2$ we sample $\bs X_g$ and $\bs \varepsilon_g$, $M=10$ times for each $g$ resulting in 10 different data sets. For each of these ten small data sets we can compute  the unpenalized softmaximin estimate (i.e. $\kappa=\infty$ in \eqref{eq21}) for a sequence of $\zeta$ values as well as the corresponding maximin estimate (i.e. $\lambda=0$ in \eqref{eq:maximin})  using base functionality in R.

\begin{figure}[H]
\begin{center}
{\includegraphics[scale=.45]{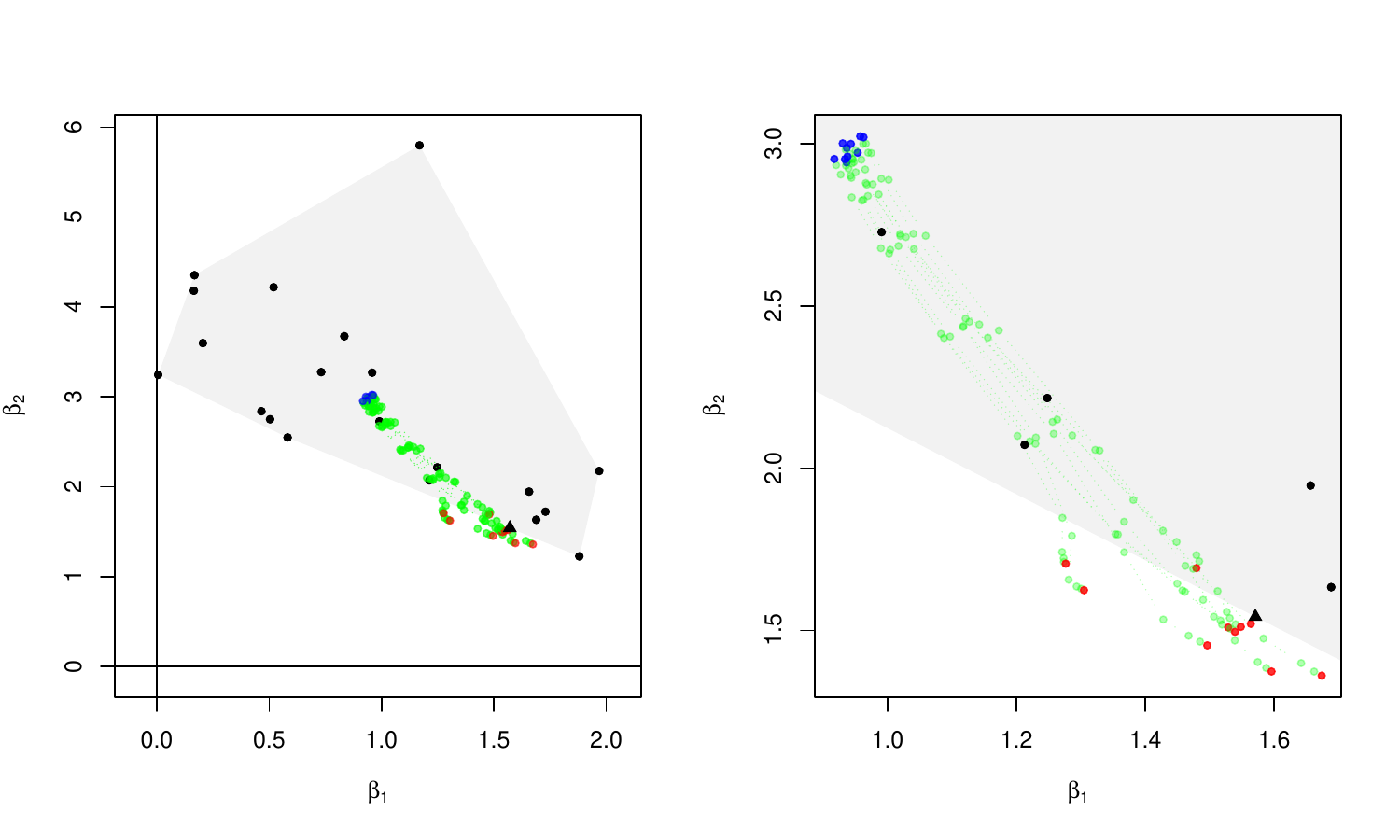}}
\caption{Left:  Convex hull (grey shaded area) of   $\supp F_B$ (black points). Theoretical maximin effect  $b^\ast$ ($\blacktriangle$), maximin estimates (red points),  soft maximin estimates (green points) for various $\zeta$,
and population LS estimates (blue points).	Right: Close up.	 			}
\label{fig:2D}
\end{center}
\end{figure}
Figure \ref{fig:2D} displays the interpolation paths of the softmaximin estimator connecting  the population LS estimates and the maximin estimates. Note that all maximin type estimates  are clustered around the theoretical maxmin effect indicated with a $\blacktriangle$ on the edge of the convex hull  of $\{b_1,\ldots,b_{20}\}$ while the pooled estimates are well inside the convex hull.

We note that  very recently  another regression method, anchor regression,  was proposed in \cite{rothenhausler2021} to handle data heterogeneity in situations where the  response distribution  can shift yielding a difference between the training data distribution and test data distribution.  If  this heterogeneity can be encoded or generated  by  a  known anchor variable their method can lead to improved and particularly more stable prediction performance. Especially by    controlling for an anchor parameter $\gamma>0$, this method   interpolates  three different  regression methods where  $\gamma=1$ yields  OLS regression and $\gamma=\infty$ an instrumental variabel regression.

 In some sense the anchor plays the role of the grouping structure $I_1,\ldots,I_G$ used in the definition of  the softmaximin estimator. However we note  that in a general setup with unknown groups, given certain structural assumptions,  we may  construct index sets $I_1,\ldots,I_G$ e.g. as in the example above or by  random  sampling.
 Theoretical guarantees in this case  are given     in \cite{meinshausen2015} for the maximin estimator and similarly to Proposition \ref{prop:zero} should extend to the soft maximin estimator.

 In addition to  mathematical complexity,  this  adds a layer of substantial computational complexity to the inference procedure since the number of groups  $G$ is then a hyperparameter that needs to be inferred  e.g. by cross validation. Thus this  clustering layer only  amplifies the importance of an efficiently computable base estimator.

\section{Computational properties}\label{sec:three}
Here we shall consider a  general  estimation setup  where  the empirical explained variance  $-\hat V_g$  from  \eqref{eq9} is replaced by a general convex group divergence function, $h_g:\RR^p\to\RR$.  Particularly, in parallel to the Bregman divergence let   $\psi:\RR^n\to \RR$ be a convex function,     and define
 \begin{alignat*}{4}
 D_\psi(x,y):=\psi(x)-\nabla \psi(y)^\top x, \quad x,y \in \RR^n.
\end{alignat*}
The   group divergence function  can then be defined  by
\begin{alignat*}{4}
 h_g(\beta):=D_\psi(\eta_g(\beta),Y_g), \quad \beta\in \RR^p,
\end{alignat*}
where $\eta_g(\beta)=\bs X_g\beta$ is the linear predictor in group $g$. Note that as $\psi$ is convex it follows that $D_\psi$, like the Bregman divergence,  is convex in its first argument and in particular $h_g$ is convex.  However, unlike the Bregman  divergence  $D_\psi$  is not non-negative.

A \emph{general} soft maximin  loss   function $ l_\zeta:\RR^p\to\RR$, is now given by
\begin{alignat*}{4}
l_\zeta(\beta):=\lse_\zeta\circ h(\beta) = \frac{\log(\sum_{j=1}^G e^{\zeta h_j(\beta)})}{\zeta}, \quad \zeta>0,
\end{alignat*}
and our  aim is to solve the    general soft maximin   problem formulated as
\begin{alignat}{4}\label{eq13}
\min_{\beta\in \RR^p}
l_\zeta(\beta) +\lambda J(\beta), \quad \lambda\geq0 .
\end{alignat}
Here $J$ is a proper convex penalty function and $\lambda $ is the penalty parameter.

Choosing $\psi$ as the square norm   yields  the negative empirical explained variance $-\hat V_g$ as group divergence i.e. $h=-\hat V$. If  $J=\Vert\cdot\Vert_1$,  both loss and  penalty are convex in this case, and \eqref{eq13}  is  equivalent to the (constrained) soft maximin problem \eqref{eq21} by strong Lagrangian duality. Hence in this case the solution to \eqref{eq13} is exactly   the soft maximin estimator.

We note that for a different choice of  $\psi$  we  would obtain an entirely new  estimator potentially with properties  very different from those of the soft maximin estimator. An immediate Mahalanobis type generalization would arise if we let $\psi$ be given as a weighted square norm. 

Solving \eqref{eq13}  in a large scale setting requires an efficient optimization algorithm for  non-differentiable  problems. 
In contrast  to the hard maximin problem,   \eqref{eq13} is, in addition to convex and non-differentiable, a (partially) differentiable and separable problem (see \cite{tseng2009}). This means  that a range of efficient algorithm will solve \eqref{eq13} e.g. first order operator splitting algorithms like ADMM or a second order algorithm like coordinate descent.
Here  we are going to consider modified versions of the proximal gradient algorithm.

\subsection{Solution algorithm}
The  proximal gradient  algorithm fundamentally works by iteratively applying the proximal operator
\begin{alignat}{4}\label{eq:4.6}
   \mathrm{prox}_{\Delta J}(\beta) = \argmin_{ \gamma \in \RR^p}
   \Big\{\frac{1}{2\Delta}\Vert \gamma - \beta \Vert_{2}^2 + J(\gamma)\Big\},\quad \Delta>0
\end{alignat}
to gradient based proposal steps. For a loss function with a Lipschitz continuous gradient
 with constant $L$, such an algorithm is guaranteed to
converge to the solution as long as  $\Delta \in (0,2/L)$, hence making it attractive to  obtain the smallest possible Lipschitz constant $L$.

With known $L$ and fixed $\Delta \in (0,2/L)$ a proximal gradient algorithm essentially consists of the following  steps:
\begin{enumerate}
\item\label{gradeval}  evaluate the gradient of the loss
\item\label{proxeval}  evaluate the proximal operator $\mathrm{prox}_{\Delta J}$
\item\label{objeval}  evaluate the loss function and penalty function.
\end{enumerate}

The computational complexity  in steps \ref{gradeval} and  \ref{objeval}
 is dominated by matrix-vector products, (see e.g. \eqref{eq9} for the soft maximin problem). The complexity in step \ref{proxeval} is  determined by $J$. As noted in \cite{beck2009} when $J$ is separable (e.g. the $\ell_1$-norm) $   \mathrm{prox}_{\Delta J}$ can be computed analytically or at low cost.

 If $L$ is not known (or if $\Delta \geq 2/L$ for a known, but perhaps
 conservative, $L$) we cannot guarantee convergence with a fixed
 choice of $\Delta$, but adding a backtracking step will ensure
 convergence of the iterates. This extra step will increase the
 per-step computational cost of the algorithm.

 When the gradient is not globally Lipschitz,  it is no longer
  guaranteed that   iterating  steps \ref{gradeval}-\ref{objeval} will
  yield a solution to \eqref{eq13} for any fixed $\Delta$. However, we  verify (Proposition \ref{prop:two}) that the following non-monotone proximal gradient (NPG) algorithm, see  \cite{wright2009} and \cite{chen2016}, will converge to a  solution of \eqref{eq13} under some   regularity conditions.

  \begin{algorithm}
 \caption{NPG minimizing $F = f + \lambda J$}
\label{alg:1}
\begin{algorithmic}[1]
\REQUIRE $\beta^0$, $L_{\max}\geq L_{\min}>0$, $\tau>1$, $c>0$, $M\in \NN $.
 \FOR{$k=0$ to $K\in \NN$}
\STATE\label{start} choose $L_k\in [L_{\min},L_{\max}]$
\STATE\label{prox} solve $\beta =\mathrm{prox}_{ \lambda J/L_k}(\beta ^{(k)}- \frac{1}{L_k}\nabla f (\beta ^{(k)}))$ \label{alg:1_3}
\IF{ $F(\beta)\leq \max_{[k- M]_+\leq i\leq k} F(\beta^{(i)})-c/2\Vert \beta-\beta^{(k)}\Vert^2$}  \label{alg:1_4}
\STATE $\beta^{(k+1)} = \beta$
\ELSE
\STATE  $L_k = \tau L_k$ and  go to \ref{prox}
\ENDIF
\ENDFOR
\end{algorithmic}
\end{algorithm}
 In particular, we show that while $l_\zeta$  does not have a Lipschitz continuous
 gradient in general,  convergence of the NPG algorithm is still guaranteed under general conditions on the group  functions $h_1,\ldots,h_G$. Furthermore, in the special case
 where $h_g = - \hat{V}_g$ with all groups sharing the same design we
 show that $l_\zeta$ has a globally Lipschitz continuous
 gradient, and we derive a Lipschitz constant.

The first result states  that $l_\zeta$  inherits strong convexity  from any  individual  group divergence function $h_g$ given that all $h_1,\ldots,h_G$ are convex and twice continuously differentiable. The proof   is given in the appendix.
\begin{thm_prop} \label{prop:one}
For $g \in \{1,\ldots,G\}$ assume $h_g$  is twice continuously differentiable and let  $w_{g,\zeta}(\beta) := e^{\zeta h_g(\beta) - \zeta l_\zeta(\beta)},\beta\in \RR^p$. Then $(w_{j,\zeta}(\beta))_{j}$ are convex weights
 and
\begin{alignat}{4}\label{eq8new}
\nabla l_\zeta(\beta)&=&&\sum_{j=1}^Gw_{j,\zeta}(\beta)\nabla  h_j(\beta)\\
 	\nabla^2 l_\zeta(\beta)
 &=&& \sum_{i=1}^G\sum_{j = i + 1}^G w_{i,\zeta}(\beta)w_{j,\zeta}(\beta) (\nabla  h_i(\beta)-\nabla  h_{j}(\beta))(\nabla  h_i(\beta)-\nabla  h_{j}(\beta))^\top\nonumber\\
 &&&+ \sum_{j=1}^G w_{j,\zeta}(\beta) \nabla^2  h_j(\beta).  \label{eq10new}
 \end{alignat}

Furthermore if  $h_1,\ldots, h_G$ are convex with  at least one $h_g$  strongly convex, then  $l_\zeta$ and  $e^{\zeta  l_\zeta}$ are strongly convex.
 \end{thm_prop}

Proposition \ref{prop:one} applies to the soft maximin loss with $h_g =-\hat{V}_g$. In this case $\nabla^2 h_g =  2\bs X^\top_g\bs X_g /n_g$, and $h_g$ is strongly convex if and only if $\bs X_g$ has rank $p$. Proposition
\ref{prop:one} implies that if one of the matrices $\bs X_g$ has rank $p$,
$l_{\zeta}$ is strongly convex. However, we also see from Proposition
\ref{prop:one} that $\nabla^2 l_\zeta(\beta)$ is not globally
bounded in general. Consider  the soft maximin loss, for instance,  with $G = 2$, $p = n_1 = n_2 = 2$ and
 \begin{alignat} {4}
\bs X_1 =
\left(\begin{array}{cc} 1 & 0 \\ 0 & 1 \end{array}\right)
\quad \textrm{and}
 \quad
 \bs X_2 = \left(\begin{array}{cc} 0 & 0 \\ \sqrt{2} &
                                                              0 \end{array}\right).
\end{alignat}

Take also $y_1=y_2=0$. When $\beta_1=\beta_2 = r\in \RR$ it holds that
$h_1(\beta) = h_2(\beta) = r^2$ and thus
$w_{1,\zeta}=w_{2,\zeta}=1/2$ for any $\zeta$, while
\begin{align*}
(\nabla  h_1(\beta)-\nabla  h_{2}(\beta))(\nabla  h_1(\beta)-\nabla
h_{2}(\beta))^\top & \\ =
\left(\begin{array}{cc} \beta_1^2 & -\beta_1\beta_2 \\ -\beta_1\beta_2
                                  & \beta_2^2 \end{array}\right)
 & =   \left(\begin{array}{cc} r^2 & -r^2 \\ -r^2 &
                                                                r^2 \end{array}\right)
\end{align*}
is unbounded.

The following result  shows, on the other hand, that for soft maximin
estimation with identical $\bs X_g$-matrices across the groups, $\nabla
l_\zeta$ is, in fact, Lipschitz continuous. The proof is in the appendix.
\begin{thm_cor}\label{coro:one}
For each $g\in\{1,\ldots,G\}$ let $\bs X_g=\bs X$  be a $m\times p$  matrix, $ \bs Y_g $ a $m\times 1$ vector and the group divergence given by
 \begin{alignat*}{4}
 h_g (\beta) = \frac{1}{m}(2\beta^\top \bs X^\top \bs Y_g-\beta^\top \bs X^\top  \bs X\beta) .
\end{alignat*}
 Then   $\nabla l_\zeta$  has Lipschitz constant $L$ bounded by
\begin{alignat}{4} \label{eq11new}
    \frac{4}{m^2}\max_{i,j}\Vert \bs X^\top (\bs Y_i-\bs Y_j) \Vert^2+ \frac{2 \vvvert \bs X^\top \bs X\vvvert }{m}\leq\frac{4\vvvert \bs X^\top\bs X\vvvert}{m^2}\Big(\max_{i,j}\Vert \bs Y_i-\bs Y_j\Vert^2+\frac{m}{2}\Big),
\end{alignat}
with $\vvvert\cdot\vvvert$  the matrix norm induced by the 2-norm $\Vert\cdot\Vert$.
\end{thm_cor}

By Corollary \ref{coro:one} if we have identical designs across groups we can obtain the soft maximin estimator by applying the  fast proximal gradient algorithm  from  \cite{beck2009}
to the optimization problem  \eqref{eq13}. Furthermore in this setting
the corollary  gives an explicit expression for the Lipschitz constant that will  yield an efficient step size $\Delta$ for the solution algorithm.

 Finally, in the general setup  the following proposition shows that   Algorithm \ref{alg:1}, which does not rely on a global Lipschitz property (\cite{chen2016}),  solves the problem \eqref{eq13}  given the assumptions  in Proposition \ref{prop:one}.
 The proof of the  proposition is given in the appendix.
\begin{thm_prop}\label{prop:two}
Assume $h_1,\ldots, h_G$ satisfy the assumptions in Proposition \ref{prop:one}. Let $(\beta^{(k)})_k$ be a sequence of iterates obtained by applying the  NPG algorithm  to \eqref{eq13}. Then   $\beta^{(k)}\to \beta^\ast$ where $\beta^\ast$ is a critical point   of $l_\zeta+\lambda J$.
\end{thm_prop}

In summary given a strongly convex group divergence function, e.g. satisfied in the soft maximin setup
when one $\bs X_g$ has full rank, we can always solve the general problem \eqref{eq13} using a proximal gradient based algorithm.

\subsection{Array tensor smoothing} \label{subsec:atsmooth}

We shall here briefly discuss an important special case, array tensor smoothing, where the design is  fixed and identical  across groups and the objective is to extract a common signal understood as a smooth function over a possibly multi-dimensional domain.  Typically the  scale of the data in this setting, will  prevent the (hard) maximin estimator from being applicable.

 By array data we mean  data with a geometry   such that it  is most naturally   organized in a multidimensional array, as opposed to  an unstructured vector. Canonical  examples are   images, movies, etc..
To formalize  this data setting and model consider a regular grid in $\RR^d$  i.e. a $d$-dimensional lattice
\begin{alignat}{4}\label{eq9new}
 \mathcal{Z}_{1}\times\mathcal{Z}_{2}\times\ldots \times \mathcal{Z}_{d}
\end{alignat}
where $ \mathcal{Z}_{j}=\{z_{ j,1},\ldots, z_{ j,m_{j}}\}\subset \mathbb{R}$ with  $ m_j\in \NN, j\in\{1,\ldots, d\}$.
Let $m:=\prod_{j=1}^dm_{j}$. If   for each group $g$ data $y_{g,1},\ldots,y_{g,m}$ is sampled  across all points in   \eqref{eq9new}
 it may be organized in a fully populated $d$-dimensional $m_{1}\times\cdots\times m_{d}$-array
 \begin{alignat}{4}\label{eq14new}
  \textbf{Y}_g=(y_{g,i_1,\ldots,i_d})_{ i_1,\ldots,i_d} , \quad i_j= 1,\ldots,m_j, \  j = 1,\ldots,d.
\end{alignat}
For this reason we refer to this type of data  as array data.

Preserving the array structure when formulating a multivariate smoothing model in this data setting   leads to the array model equation
\begin{alignat}{4}\label{eq6}
  {Y}_{g, i_1,\ldots,i_d}= f_g(z_{1,i_1},\ldots,z_{d, i_d}) + \epsilon_{g,  i_1,\ldots,i_d}, \quad z_{ j,i_j}\in
  \mathcal{Z}_j,
\end{alignat}
for each $g\in \{1, \ldots,G\}$, where $f_g$ is a  smooth group  signal and $\epsilon_{g,  i_1,\ldots,i_d}$ an appropriate error term.

To obtain a linear array model from \eqref{eq6} we  parameterize  $f_g $  using a basis expansion. A particularly convenient way of representing a multivariate function, is to use the tensor product construction to specify   multivariate basis functions  in terms of  (tensor) products of families of  univariate  basis functions  $((\varphi_{j,k})_{k=1}^\infty)_{j=1}^d$. That is with   $\varphi_{j,k} : \mathbb{R} \to \mathbb{R}$  the $k$th basis function in the $j$th dimension  ($k$th $j$-marginal basis function),
 we    can represent the smooth signal as
 \begin{alignat}{4}\label{eq7}
f_g(z_1,\ldots,z_d)= \sum_{k_1,\ldots,k_d} \Theta_{g,k_1,\ldots,k_d}\prod_{j=1}^d\varphi_{j,k_j}(z_j), \quad
z_j\in \RR,
\end{alignat}
 where $ (\Theta_{g,k_1,\ldots,k_d})_{k_1,\ldots,k_d}$ are  basis coefficients.

Now, in order to implement this representation for the model \eqref{eq6}  we need to determine the number of basis functions to use in the $j$the dimension, $p_j\in \NN$,  to obtain an  (finite)  approximation of $f_g$.
For tensor product basis functions it is customary to choose $p_j$ as a function  of the cardinality of $\mathcal Z_j$   e.g. $p_j=[m_{j}/5]$ (see \cite{currie2006}).

 With $p_j\in \NN$ fixed for each $j\in\{1,\ldots, d\}$ we obtain the model \eqref{eq6} as a linear array model   in the following way.  For  each $j\in \{1,\ldots, d\}$ define a $m_j\times p_j$ matrix  $\Phi_j = (\varphi_{j,k}(z_{j, i}))_{i, k}$  containing the   values of the $p_j$ basis functions   evaluated  at  the $m_j$ points in $\mathcal Z_j$.  We call $\Phi_j$ a marginal design matrix. Also define a   $p_1\times \cdots\times p_d$-array $\bs{\Theta}_g=(\Theta_{g,j_1,\ldots,j_d})_{j_1=1,\ldots,j_d=1}^{p_1,\ldots,p_d}$  containing      the corresponding  basis coefficients.

  It then follows directly from   the identity  \eqref{eq7}  that the tensor (Kronecker) product of these marginal design matrices,
\begin{alignat}{4}\label{eq15}
  \Phi =   \Phi_{d}\otimes \cdots\otimes \Phi_{2} \otimes \Phi_{1},
\end{alignat}
is  the     design matrix  for the linear model version of  \eqref{eq6}. This means that  we can in principle implement   \eqref{eq6} as a standard linear model using $\Phi$. However   from a computational complexity perspective that  is suboptimal and potentially not feasible   in large scale data settings as $\Phi$ grows as $\prod p_j\prod m_j$.

Instead,  we can exploit the array tensor structure of the problem and   only rely on  the much smaller marginal matrices. As  shown in \cite{currie2006} any linear model with   array structured  data \eqref{eq14new} and  tensor structured design \eqref{eq15} can be formulated as a   linear array model and fitted using   so-called array arithmetic. The key computation is the   rotated $H$-transform $\rho$, (see  \cite{currie2006}  for details),  that allows us to write the model \eqref{eq6}  as a linear array model
 \begin{alignat}{4}\label{eq12}
  \textbf{Y}_g=\rho(\Phi_{d},\rho(\Phi_{d-1},\ldots, \rho(\Phi_{1}, \bs{\Theta}_g))) + \bs{\epsilon}_g,
\end{alignat}
where  $\bs{\epsilon}_g$ is a $m_1\times\cdots\times m_d$
array containing  the error terms.

As indicated by \eqref{eq12}, using $\rho$  the design matrix-parameter vector products,  needed  in steps \ref{gradeval} and \ref{objeval} above, are computed  without having access to the (very large) matrix $\Phi$. In addition the computation has lower complexity than the corresponding matrix-vector product (\cite{deboor1979}, \cite{buis1996}).

Finally, the tensor structure in \eqref{eq15} makes the constant $L$  in Corollary \ref{coro:one}  easy to compute, see (30) in \cite{lund2017a}. The  implication is that we can run the proximal gradient algorithm without performing any backtracking.

Following \cite{lund2017a} we have implemented
both the fast proximal algorithm as well as the NPG algorithm  \ref{alg:1}
in a way that exploits the array-tensor structure  described above. Implementations are provided for 1D, 2D, and 3D array data in the R package \verb+SMME+,  \cite{lund2021cran} along with the  implementation for general unstructured data.

\section{Numerical experiments}\label{sec:four}
To demonstrate the properties of the soft maximin  estimator we   present two data examples. 
The first example  is a  real data set,  also analyzed in \cite{rothenhausler2021}, with seemingly a   high signal to noise ratio and   moderate  data heterogeneity.  We use this data set to highlight the  interpolation property  inherent in the soft maximin methodology.

In the second example we use a simulated large-scale data  set with  low signal to noise ratio and strong heterogeneity to benchmark our methodology.
We compare the run time and prediction accuracy of the soft maximin estimator to that of the pooled OLS estimator  and  the maximin aggregation method, see \cite{buhlmann2016}. This simulation  based example is inspired by   a large scale neuronal data set analyzed  in Appendix \ref{sec:neuron}.

\subsection{Washington DC bike sharing data} \label{sec:bike}
\begin{figure}[H]
\begin{center}
\includegraphics[scale=0.47]{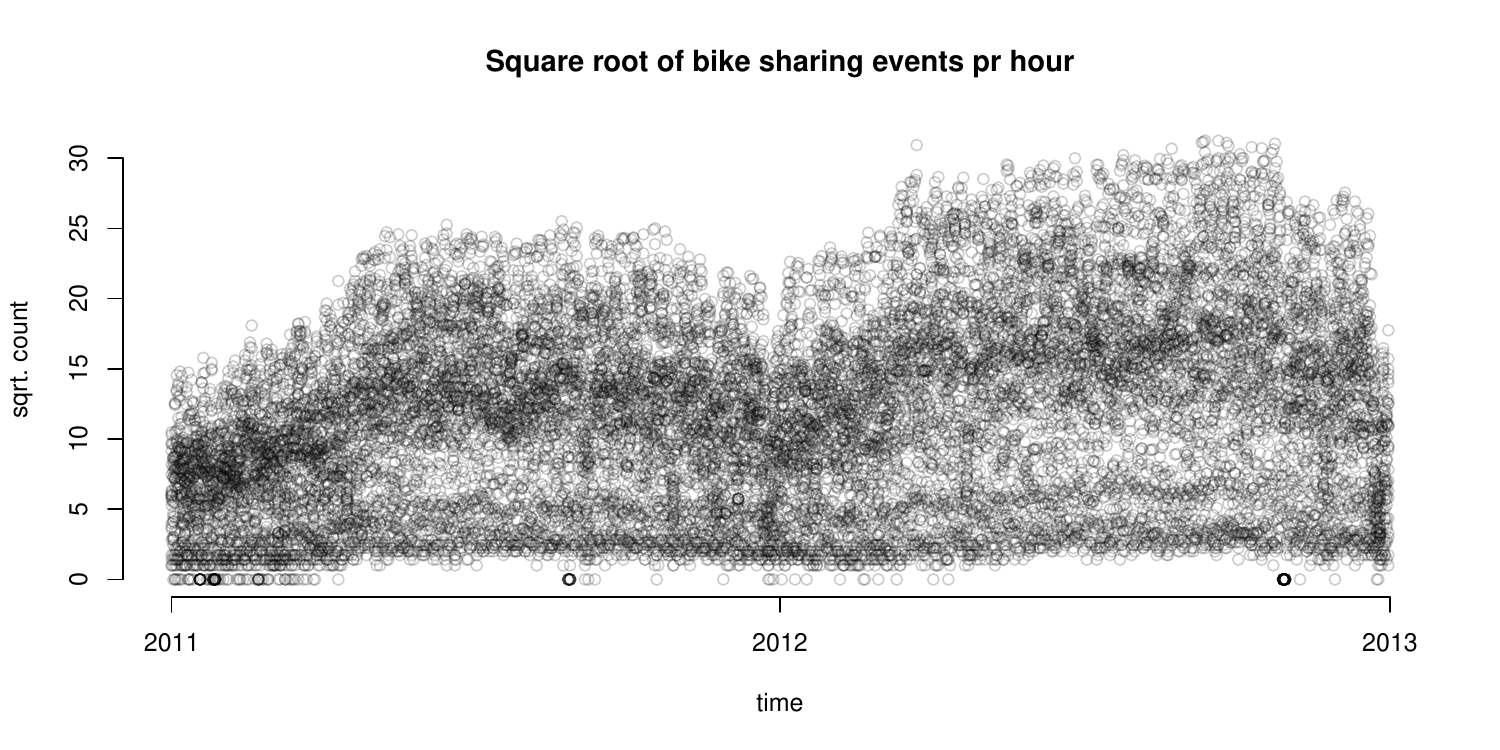}
\caption{Hourly number of bike shares in Washington DC 2011 and 2012.}
\label{fig:3newnew}
\end{center}
\end{figure}

The data used to  produce the results in Figure \ref{fig:1new} are described in \cite{fanaee-t2013}. The data set contains two years (2011 and 2012) of data (variable \verb+cnt+ see Figure \ref{fig:3newnew}) from a bike sharing scheme in Washington DC along with auxiliary data presumably  relevant for bike usage e.g. weather.   We  model  the hourly number of bike shares from the data set shown  in Figure \ref{fig:3newnew}. 
Specifically   the square root of the number of bike events (\verb+cnt+) is modelled as a smooth function of hour of day (\verb+hr+ with values 0 to 23), smooth function of day of week (\verb+weekday+ with values 0 to 6) and the weather situation (\verb+weathersit+ with levels 1,2,3). The model equation for observation $i$ can be written as
 \begin{alignat}{4}\label{eq20new}
\sqrt{\texttt{cnt}_i}=\sum_{j=1}^{10}\alpha_j \phi_j(\texttt{hr}_i)+\sum_{j=1}^{5}\beta_j \phi_j(\texttt{weekday}_i)+ \sum_{j=1}^3 \gamma_j 1_j(\texttt{weathersit}_i) +\epsilon_i
\end{alignat}
where $\phi$ are cubic basis spline functions.

Note the raw data do not contain any hours with zero counts. However as there are 165  hours unaccounted for in the data set, we suspect these  are hours with zero counts (e.g. during hurricane Sandy in Oct. 2012) and  impute this data for simplicity. This imputation has no effect on the analysis.  Also there are three observations that have \verb+weathersit+ = 4. To have all  \verb+weathersit+ levels   present for the relevant test and train split we change the level of these  observations to 3. Again this change has no effect relative to leaving out the observations entirely.

The data do not a priori have a  grouping structure that  explains the heterogeneity. However,   a safeguard against temporal heterogeneity may be  obtained by  using the variable \verb+mnth+  to group the data in months.  Using this grouping, to show the effect of the temporal  heterogeneity,  we perform a simple experiment where we i)    fit the model on 2011 data and predict on 2012 data and    ii) fit the model on 2012 data and predict on 2011 data.

We fit the model \eqref{eq20new} to data using the unpenalized soft maximin estimator ($\lambda=0$  in \eqref{eq13}) and  the pooled OLS estimator. We also fit an extension of \eqref{eq20new} that includes a smooth function of  the grouping variable \verb+mnth+ using OLS. This model potentially controls for the heterogeneity and might have superior performance. Note the aim of the analysis is not to obtain the model with best predictive performance but rather to highlight how maximin type estimators can safeguard against  heterogeneity. In  settings where the source of structured variation is more opaque and cannot be taken into account easily in the model this idea may still work.

Figure \ref{fig:1new}  shows  the cumulative root mean squared error 
\begin{alignat}{4}\label{eq21new}
\text{RMSE}_\zeta(t)=\sqrt{\frac{1}{t}\sum_{i=1}^t (y_i - \hat y_{i,\zeta})^2}
\end{alignat}
on both training data and test data for the model fit for respectively i) and ii). Here  $y=\sqrt{\texttt{cnt}}$ and $\hat y_{\zeta}$, the fitted values  for resp. pooling ($\zeta=0$) and soft maximin  ($\zeta>0$),  are chronologically ordered.
 \begin{figure}[H]
\begin{center}
\includegraphics[scale=0.39]{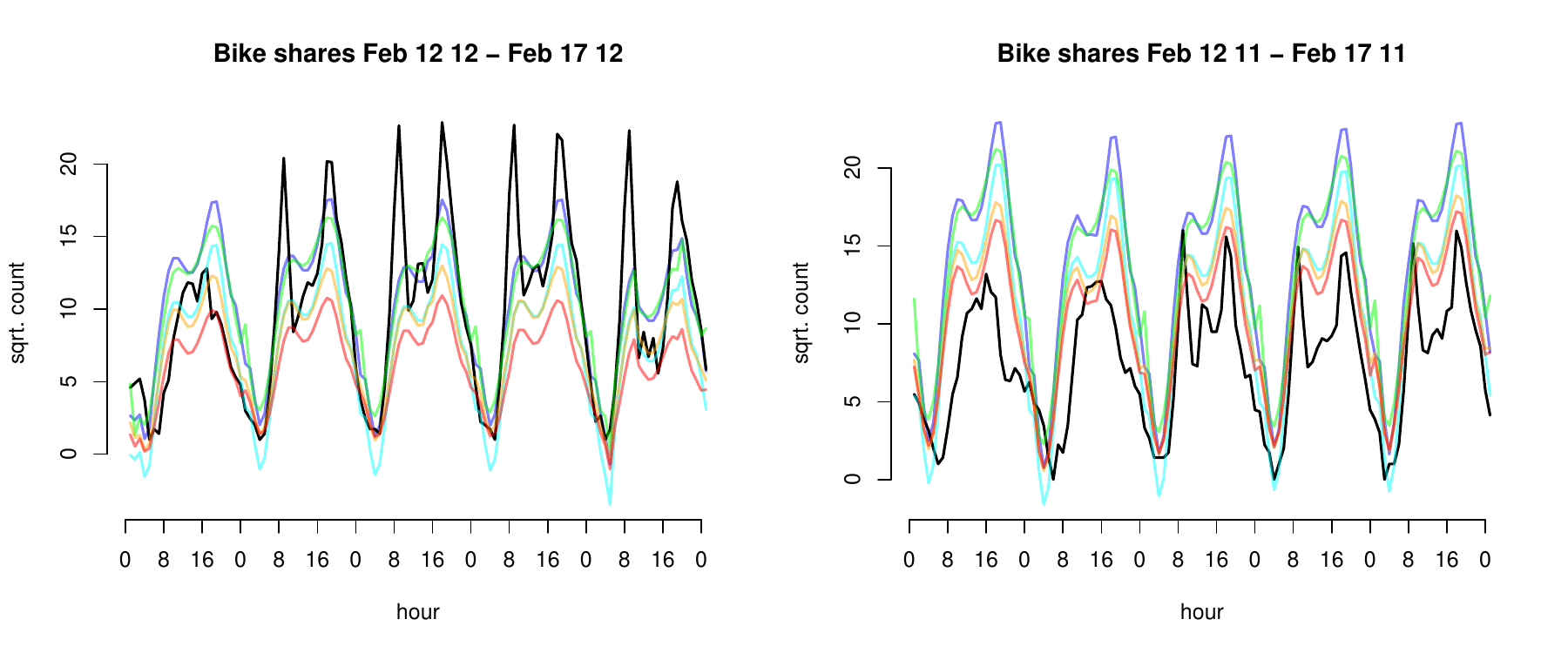}
\includegraphics[scale=0.46]{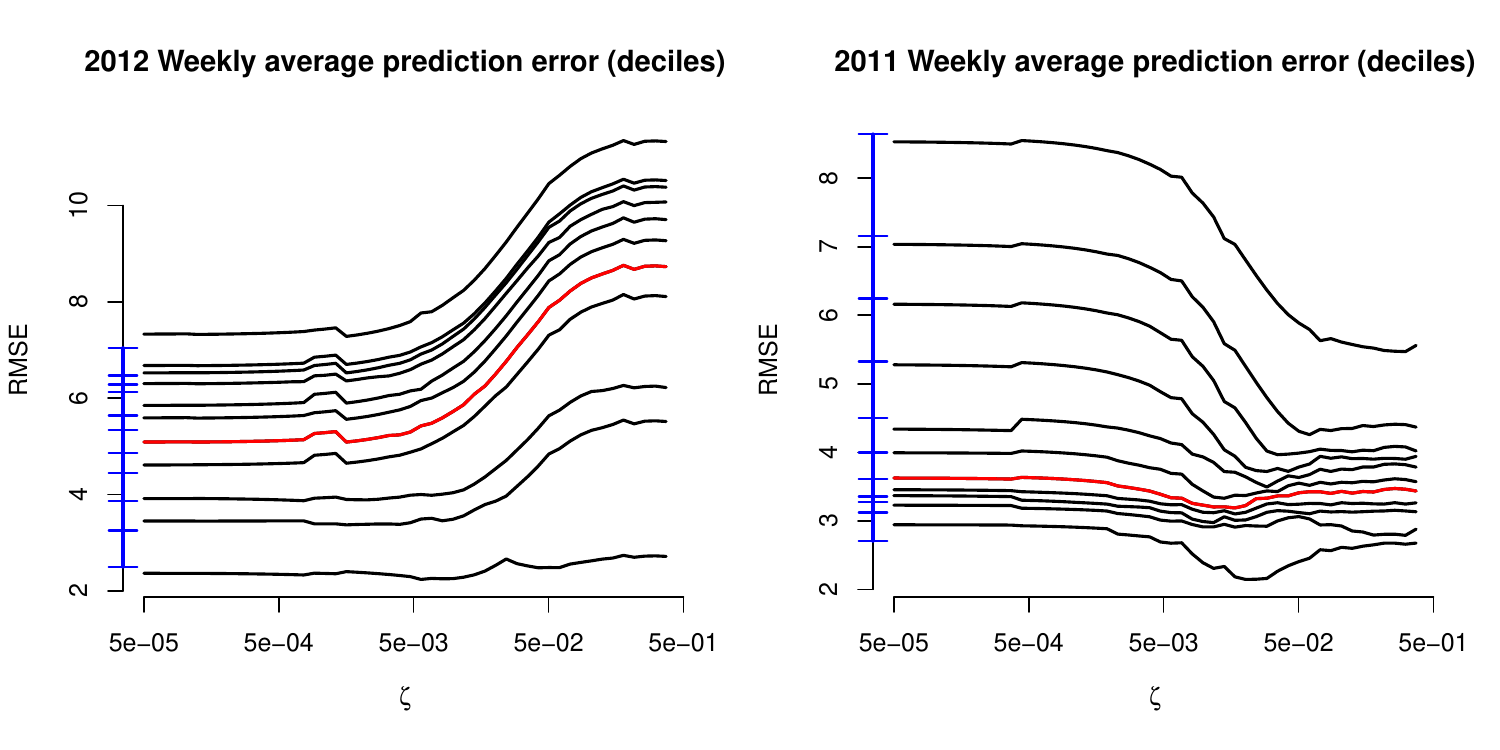}
\caption{Top: Predictions for 2012  when training on  2011 (left) and for 2011     when training on 2012 (right),  for the soft maximin estimator with $\zeta\in\{0.0001, 0.01,1\}$ (resp. green, orange and red), the pooled OLS estimator (blue and cyan) and   $\sqrt{\texttt{cnt}}$ (black).  Bottom:  Deciles of  weekly averaged prediction error  on 2012  data  for 2011 training data (left) and on 2011  data  for 2012 training data (right) for  soft maximin estimation as a function of $\zeta$. Blue part of the $y$-axis shows deciles of the pooled estimator. Red curve indicates the median.}
\label{fig:4new}
\end{center}
\end{figure}
The findings in Figure \ref{fig:1new} are in line with Figure \ref{fig:4new} that shows the predictions (out of sample) for 5 consecutive days   for  experiments i) and ii).  In the left panel we observe  that the high $\zeta$ soft maximin (red) underfits the 2012 data while the low  $\zeta$ soft maximin like pooling predicts well. Conversely in the right panel the low $\zeta$ and pooling overfits the high levels observed in some parts of the 2012 data and the result is worse predictions on the 2011 data.

The bottom panel in Figure \ref{fig:4new} shows the deciles of  RMSEs for the test set (2012 resp. 2011) averaged for each week as a function of $\zeta$. Training on 2011 data and testing on 2012 data 
 shows  weekly averaged prediction error  of the pooling estimator  to have lower median and more narrow support than soft maximin estimators. Conversely   training on 2012 data and testing on 2011 a strictly positive $\zeta $ yields more stable predictions and also a lower median. Note the Figure resembles that of Figure 5 in \cite{rothenhausler2021}.

In   Appendix \ref{sec:bike_app}  we fit alternative models that include temperature and humidity. The results are similar to those presented above.  We also fit a model where we use  \verb+weathersit+ as grouping.  For this grouping the soft maximin estimators still appear  attractive   but the picture is less clear.

Also in the appendix   a cross validation scheme is used  to tune   $\zeta$ for experiment i) and ii). The results are in line with the findings in figures \ref{fig:1new} and \ref{fig:4new} and suggest that  hard maximin is not optimal in neither i) nor ii). For i)  pooling ( $\zeta=0$) works best and for ii) a $\zeta$ around 0.03  results in the lowest prediction error. Fitting a spectrum of estimators can  be advantageous compared to fitting either of the extreme estimators, pooling respectively maximin,  in the given context.

\subsection{Benchmark on simulated array data} 
To benchmark  the soft maximin method against existing alternatives we set  up   a prediction experiment for the model described in Section \ref{subsec:atsmooth} on a simulated data set.   In appendix \ref{sec:neuron} we carry out the same experiment on a real large scale neuroscientific data set. The  experiment is  structured like $K$-fold cross validation aimed at discerning the optimal choice of hyperparameters $\zeta$ and $\lambda$ in a grid search. 

To evaluate   the performance of the soft maximin estimator as a function of the parameter $\zeta$ we train the  model for  $\zeta\in\{2, 100,  200\}$. We also compute the pooled estimator corresponding to $\zeta=0$ and the maximin aggregation (magging) estimator from \cite{buhlmann2016}. In general magging is approximately (hard) maximin estimation and should therefore correspond  to   $\zeta=\infty$. 

This in turn entails  solving five  different $\ell_1$-penalized estimation problems:
\begin{itemize}
\item To obtain the  three soft maximin estimators we need to solve the  problem \eqref{eq13}  with  $\ell_1$-norm penalty for each $\zeta \in\{2, 100, 200\}$. To do this  we use the R package  \verb+SMME+, \cite{lund2021cran}.
\item  With identical (fixed) design across groups, we  obtain the (penalized) pooled estimator as  (penalized) regression of the empirical average across groups on the fixed design. We  use the R package  \verb+glamlasso+, \cite{lund2018a}  to solve the resulting lasso problem.

\item For the  magging estimator  we  have to  solve a lasso problem for each group, given $\lambda $ and the design.  We use the R package  \verb+glamlasso+, \cite{lund2018a} to obtain the individual group fits. These fits are then  maximin aggregated (magging)  across groups  by  solving a quadratic optimization problem as proposed in    \cite{buhlmann2016}.
\end{itemize}

All computations are carried out on a Macbook Pro with a 2.8 GHz Intel core i7
processor and 16 GB of 1600 MHz DDR3 memory.

\subsubsection{Simulated array data}\label{subsec:3dim}
 \begin{figure}[H]
\centering
\includegraphics[scale=0.65]{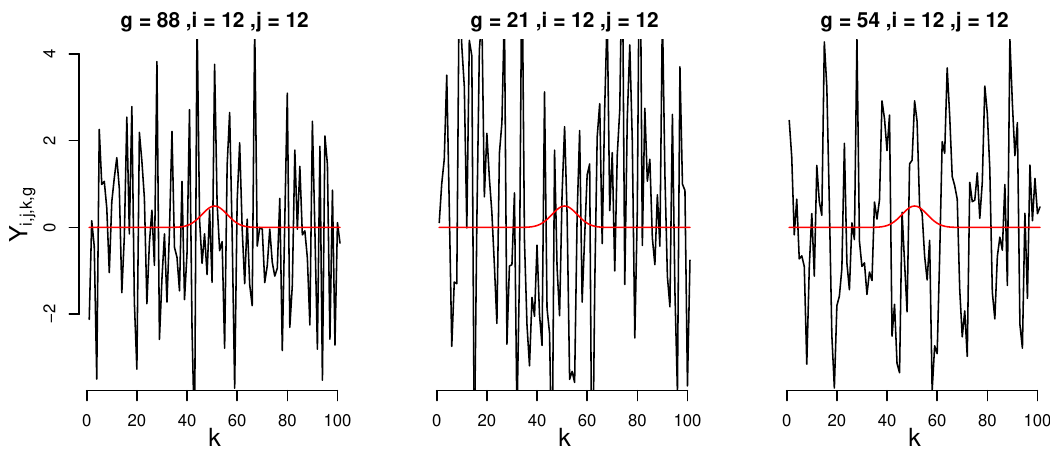}
\caption{Temporal plot of the simulated data (black) for three different groups and the underlying common Gaussian signal (red). }
\label{fig:two}
\end{figure}
 We simulate  data with three components: i) a common Gaussian signal of interest $\phi(x,y,t)=200\phi_{12.5,4}(x)\phi_{12.5,4}(y)\phi_{50,25}(t)$   ($\phi_{\mu,\sigma^2}$ is the density for the $\mathcal{N}(\mu,
\sigma^2)$ distribution) superimposed with
ii) periodic group specific signals with randomly varying frequency and phase and
iii) additive white noise.
Specifically for each $g\in\{1,\ldots,G\}$ the 3-dimensional  data array  was simulated  according to
\begin{alignat}{4}\label{eq19new}
  Y_{g,i,j,k}&=\phi(x_i,y_j,t_k)
  +5 \sum_{j\in J_g} \varphi_j  (x_i + p_g)\varphi_j  (y_i + p_g)\varphi_j  (t_k + p_g)+\epsilon_{g,i,j,k},
\end{alignat}
 with $x_i=1,2,\ldots,25$, $y_i=1,2,\ldots,25$ and
$t_i=1,2,\ldots,101$.
Here $J_g$ is a set of $7$ integers sampled uniformly from  $ \{1,\ldots,101\}
$, $\varphi_j$ is the $j$th Fourier basis function, $p_g\sim
\mathrm{unif}(-\pi,\pi)$, and $\epsilon_{g,i,j,k}\sim
\mathcal{N}(0,10)$.

 We note that the common 3-dimensional Gaussian signal $\phi$, due to its light tails,  is spatially as well as temporally localized.

  Figure \ref{fig:two} shows the simulated signals for three different groups plotted across time for $(i,j)=(12,12)$.  The common signal is  dominated by  group specific fluctuations in each group and not visually apparent.

\subsubsection{Experiment setup}
We use the array-tensor model from  Section \ref{subsec:atsmooth} with  B-splines   as basis functions in each dimension. The   number of basis functions used in each dimension is respectively, $p_1=p_2=10$ (spatial) and  $p_3=23$ (temporal). This gives us  an array model with marginal design matrices  $\Phi_1$, $\Phi_2$, and $\Phi_3$, of size $25\times 10,25\times 10$, and $101\times 23$ respectively, given by  B-spline  function evaluations over the marginal domains.  The model has  $p=2300$ parameters.

To set up the experiment we simulate  $G = 100$ groups of  3-dimensional signals according to \eqref{eq19new}.  We then randomly sample $K=7$ folds (14 groups in each fold). This gives us a total of $m=883,750$ observations in each fold.

For each method we train   the model  on each fold and test on the remaining 6 folds. By repeating this  procedure $N=10$  times we obtain 70 fits  and corresponding test metrics for each method  and each setting of $\lambda$.

\subsubsection{Experimental results}
\begin{figure}[H]
\centering
\includegraphics[scale=0.5]{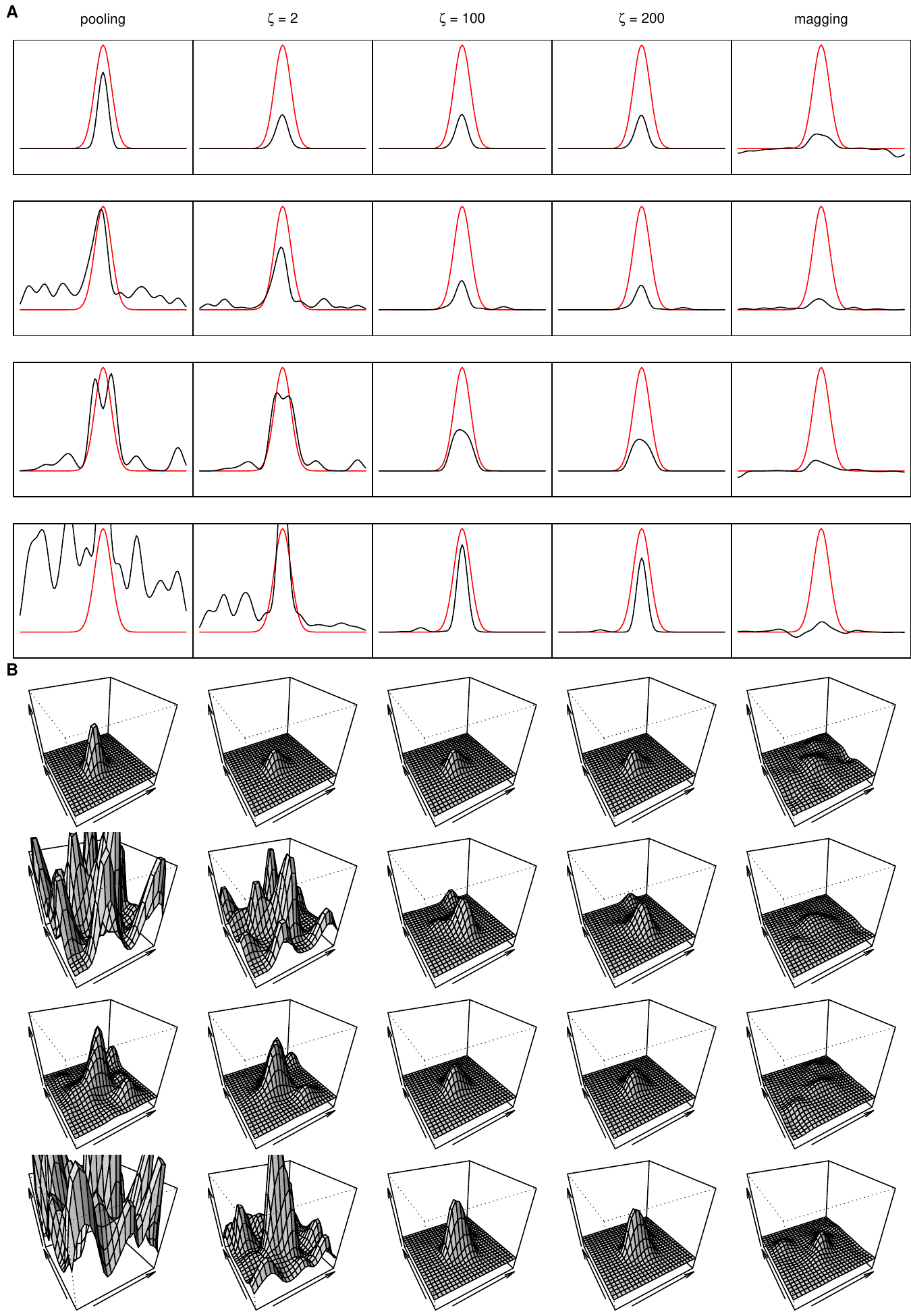}
\caption{A: Temporal plots  for  $(x,y)=(12,12,)$.   True signal $\phi(x,y,t)$  (red) and  estimate $\hat Y_{x,y,t}$ (black) for model no. 10.  Columns left to right; pooling, soft maximin $\zeta\in\{2,100, 200\}$  and magging. Each row corresponds to a  training set with  14 groups. B:  Spatial plots for $t=50$ of the estimates   from panel A.
}
\label{fig:1}
\end{figure}

Predictions for  the experiment for 4 different validation sets are shown  in Figure \ref{fig:1}. For one data set the pooled estimator is succesful in extrating a clear Gaussian signal but it fails on the other three. In contrast the soft maximin estimators ($\zeta\in \{100,200\}$) perform well on all sets and display less variation.

\begin{figure}[H]
\begin{center}
\includegraphics[scale=0.4]{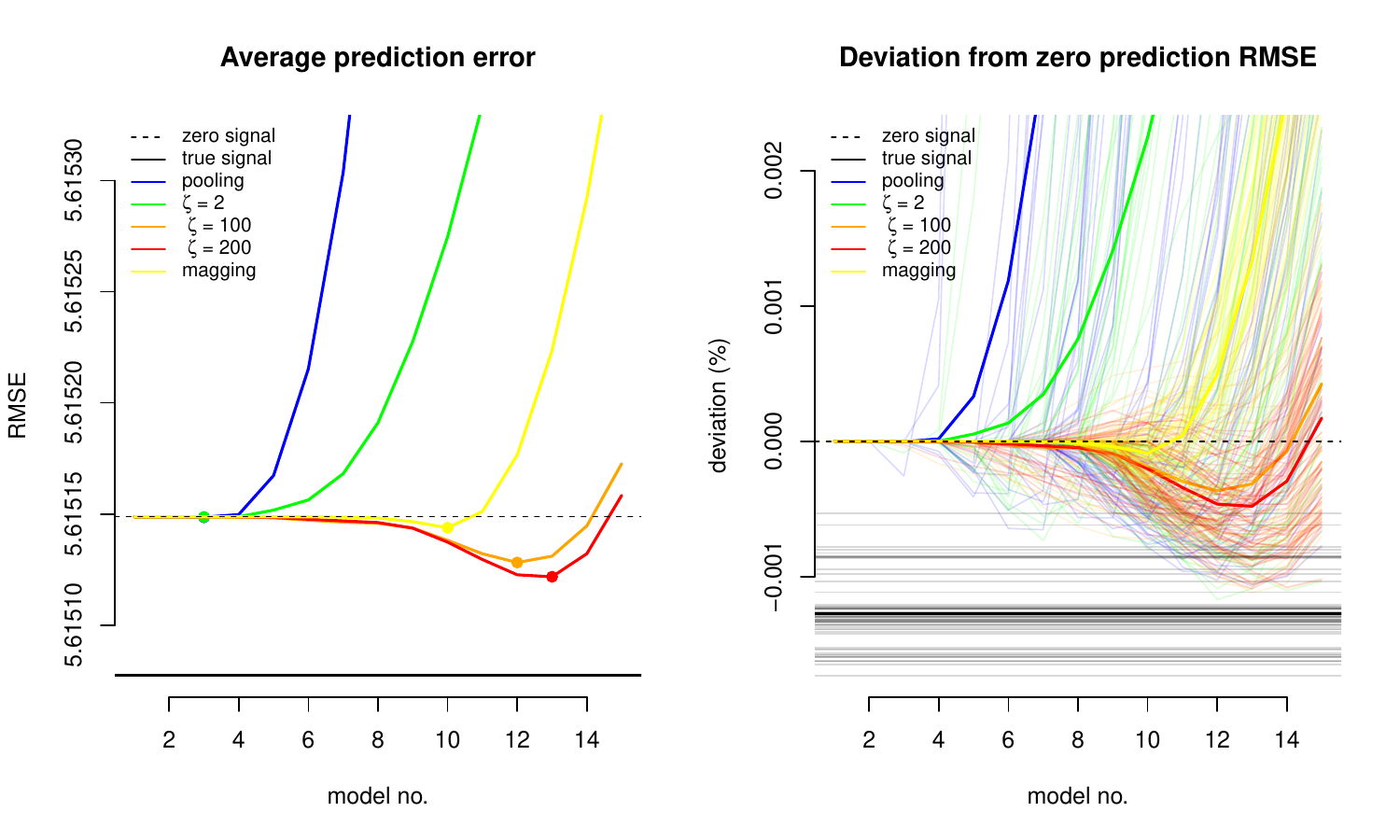}
\caption{Left: Average RMSE across  70  test sets as a function of model complexity (model no.).
 Zero signal (dashed), true signal (black), pooled estimator  (blue), soft maximin   $\zeta=2$ (green), soft maximin   $\zeta=100$ (orange), soft maximin   $\zeta=200$ (red)  and magging (yellow). The minimum is indicated with a bullet.
 Right:  Deviation  in RMSE relative to  the zero prediction RMSE (percentage difference) for each method and on each test set (thin lines), as a function of model complexity. The corresponding averages are indicated with a thick line of the same color.}
\label{fig:3new}
\end{center}
\end{figure}

Figure \ref{fig:3new} shows the average of the root mean squared prediction error (RMSPE)   as a function of model complexity ($\lambda$) for each method. RMSPE is defined as  $\Vert \hat Y_{x,s}-Y_{-s}\Vert_2/\sqrt{m}$ where $ \hat Y_{x,s}$ is the prediction from training on set $s$ using method $x$ and $Y_{-s}$ are the observations in the complement to $s$. 
 
   The dashed line represents the average error made when using the zero signal,  the most conservative estimate. The black line on the other hand is the average error when using the true signal as predicted values and is the optimal prediction for this data. We see the best performing method is the soft maximin with $\zeta=200$ (red) while the worst is the pooled estimator (blue). In particular, the pooled estimator performs no better than the zero prediction. Surprisingly the magging estimator (yellow) does not perform  much better than the zero prediction on average.

The right display in Figure \ref{fig:3new} illustrates the variability in prediction accuracy for the different methods using their relative deviation in RMSE  from that of the zero prediction.  The low $\zeta$ estimators (pooled and $\zeta =2$) display high variability, reflecting a tendency to overfit group specific signals in the  data.  The high $\zeta$ soft maximin estimators and magging  show much less variability. This underlines the robustness of the estimation methodology.

We note that only the high $\zeta$ methods succeed in extracting a  common signal that is significantly more accurate  than the zero prediction.

We note that while the gain in prediction performance,  relative to the zero prediction is  small, due to the low signal to noise ratio, it is not insignificant in terms of the quality of the extracted signal as illustrated in Figure \ref{fig:1}.  Here, the fit on four different training sets are displayed,  and we  observe directly  how the low $\zeta$ methods tend to fit group fluctuation compared to the high $\zeta$ methods.

\begin{figure}[H]
	\centering
	\includegraphics[scale=0.4]{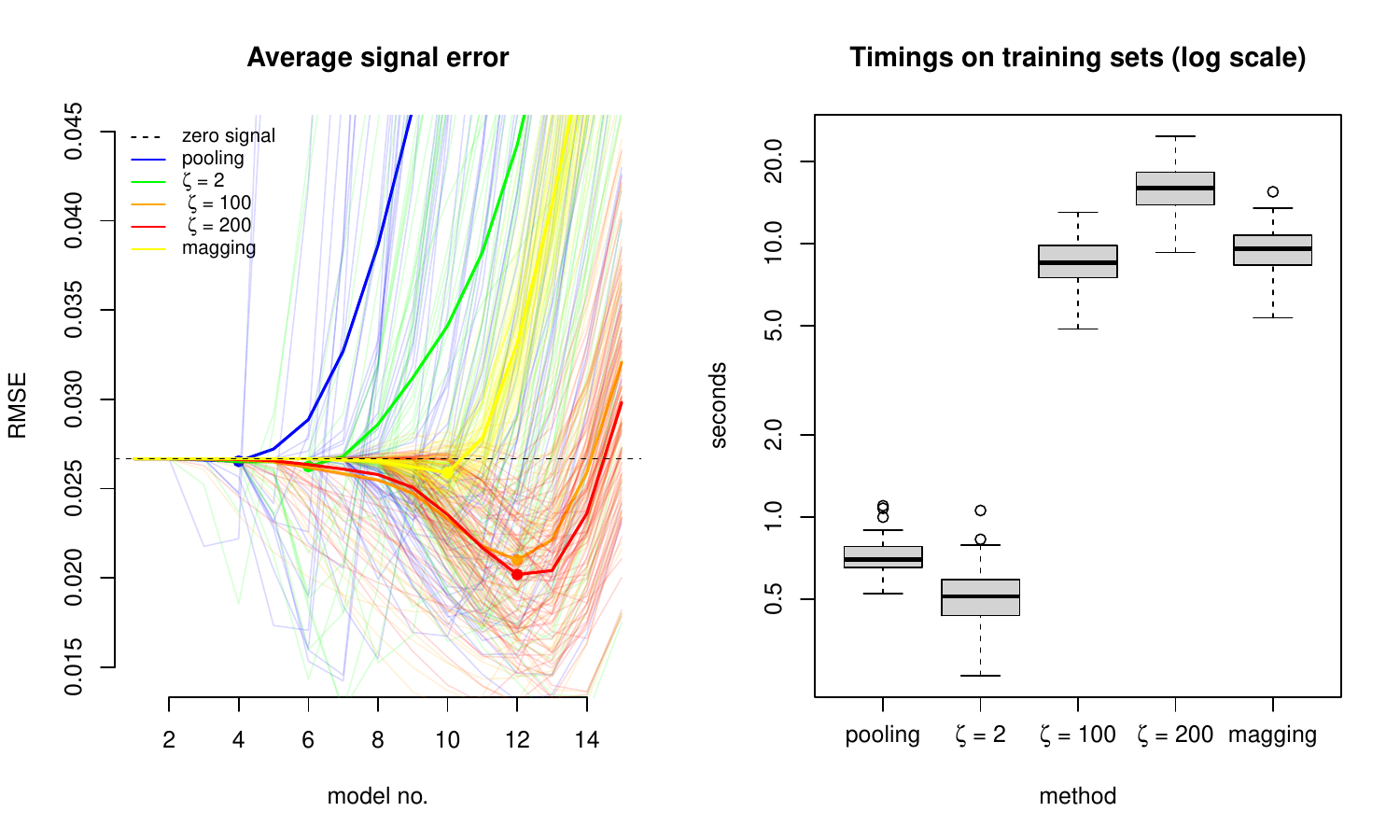}
	\caption{Left:  mean squared signal error based on each of the 70 training set. Pooled estimator  (blue), soft maximin   $\zeta=2$ (green), soft maximin   $\zeta=100$ (orange), soft maximin   $\zeta=200$ (red)  and magging (yellow). Right: Summary of run times (log scale) for the 70 training sets for each method.}
	\label{fig:4}
\end{figure}

We   quantify this   by computing the average of the root mean squared signal error $\Vert \hat Y_{x,s} - \phi\Vert_2/\sqrt{m}$ made by the prediction obtained by training  on set $s$ using method $x$.  The left display in Figure \ref{fig:4} shows the result for each method $x$  and set $s$ as a function of model complexity as well as the average over $s$ (folds) and  confirms the impression from Figure \ref{fig:1} and Figure \ref{fig:3new}.

Finally, the right display in Figure \ref{fig:4} shows how  each method performs in terms of  run time. In this setting given the average response across  the 14 groups in a fold, computing the pooled estimator  has the same complexity as computing one group fit in the magging procedure. Not surprisingly the pooled estimator (0.8 s) is roughly 12 times faster than magging (9.7 s), and also faster than the high $\zeta$ methods (8.1 s resp. 16.6 s). However,  the soft maximin estimator with  $\zeta=2$ (0.5 s) is faster than the pooled least square estimator.

\section{Discussion}\label{sec:five}
The maximin estimator with the $\ell_1$-penalty, as defined in
\cite{meinshausen2015}, solves the minimization problem
 \begin{alignat}{4}
\label{eq:maximin}
\hat\beta_{mm}:=\arg\min_{\beta} \max_g\{ - \hat{V}_g(\beta) \} + \lambda \| \beta \|_1.
\end{alignat}
Though the objective function is convex, it is nondifferentiable as
well as nonseparable, and
contrary to the claim in Section 4 of \cite{meinshausen2015},
coordinate descent will not always solve \eqref{eq:maximin}, see \cite{tseng2009}.

Two approximate  approaches for solving \eqref{eq:maximin}
were suggested in  \cite{meinshausen2015}. The first, also a
 smooth approximation of the term $\max_g \{ -\hat{V}_g(\beta)\}$, however,  appears theoretically invalid and we did not find it to work in practice either. The second approximation, the maximal penalty solution, obtains a solution to \eqref{eq:maximin} for the maximum $\lambda$ that yields a non-zero solution (at least one active feature) to \eqref{eq:maximin}. This solution is appropriate as an efficient initial estimator and clearly much cruder than a finely tuned maximin (type) estimator.

We note in passing that the  solution path of \eqref{eq:maximin} is piecewise
linear in $\lambda$, and it may thus be computed using a  method like LARS, see
\cite{roll2008}. A LARS-type algorithm or a coordinate descent
algorithm of a smooth majorant, such as the soft maximin loss, has subsequently also been   proposed
to us by Meinshausen (personal communication) as better
alternatives to those suggested in \cite{meinshausen2015}. In our
experience, the LARS-type algorithm scales poorly with the size of the
problem, and neither LARS nor coordinate descent can exploit
the array-tensor structure.

 We have developed soft maximin estimation as an alternative to maximin estimation that retains desirable statistical properties
and is computationally more efficient. Furthermore  the soft maximin parameter $\zeta$  controls the tradeoff between groups with large explained variance and
groups with small explained variance leading to an interpolation of pooled estimation and maximin estimation. The gradient representation \eqref{eq8new} shows explicitly how this tradeoff works: the gradient of the soft maximin loss is a convex combination of the
gradients of the group-wise   loss functions with weights controlled by
$\zeta$. The largest weights are on those groups with the smallest
explained variances and as $\zeta \to \infty$ the weights concentrate on the groups with minimal explained variance.

 On the bike sharing data we have illustrated this interpolation property and the benefit of having a spectrum of estimators available instead of  only the extremes (pooling resp. maximin). Notably, the optimal value of $\zeta$   in terms of prediction accuracy depends on the data context. Specifically we conclude that for heterogeneous data  (hard) maximin estimator is  not necessarily the best choice. Instead by  tuning $\zeta$ e.g. by cross validation it might be possible to obtain  an estimator in the spectrum between hard maximin and pooling that works better for the specific context. We note that a similar interpolation idea   for heterogenous data has recently been proposed  in the so called  anchor regression framework, see \cite{rothenhausler2021}. 
 
 On  the simulated data (see also the neuronal  VSDI data in   Appendix \ref{sec:neuron})  it was demonstrated how the soft maximin estimator  was able to extract a signal, and how the choice of the tuning parameter $\zeta$ affects the extracted signal and the prediction performance. In particular  the simulations showed that soft maximin estimation ($\zeta\in\{100,200\}$) was able to extract a signal even in the presence of large heterogeneous noise components where the other methods (pooling and magging) failed. 
 
In addition,  we note that magging,  proposed in \cite{buhlmann2016} as  a computationally attractive and generic alternative
to \eqref{eq:maximin} for estimation of maximin effects,   in our numerical experiment, is not    faster than using the soft maximin estimator. 

In summary  our proposed algorithm provides a
means for approximately minimizing \eqref{eq:maximin} and is as such an
alternative to magging as an estimator of the maximin effect. More importantly, by the introduction of the tuning
parameter $\zeta$ in the soft maximin loss we not only achieved an
approximate solution of  \eqref{eq:maximin} but an interpolation
between the (hard) maximin estimator and the pooled WLS estimator.

We expect that soft maximin estimation  will be practically useful in a number of different
contexts,  as a way of aggregating explained variances
across groups. In particular because it down-weights groups with a
large explained variance that
might simply be outliers, while it
does not go to the extreme of the maximin effect, that can kill the signal
completely.

\appendix
\section{Proofs}\label{app:a}
\begin{proof}[Proof of Lemma \ref{lemma:0}]
i)  Since $x_g= \max\{x_1,\ldots,x_G\}$ for some $g\in\{1,\ldots,G\}$,
\begin{alignat*}{4}
 \max\{x_1,\ldots,x_G\}=  \frac{\log(e^{ \zeta x_g} )}{\zeta}
 \leq   \frac{\log(e^{\zeta x_g} +\sum_{j\neq g}e^{ \zeta x_j})}{\zeta}=\lse(x)
\end{alignat*}
and also
\begin{alignat*}{4}
  \lse(x)\leq \frac {\log(\sum_j  e^{\zeta x_g} )}{\zeta}=
   \frac {\log(G  e^{\zeta x_g} )}{\zeta}=
   \frac {\log(G  )}{\zeta}+ \max\{x_1,\ldots,x_G\},
\end{alignat*}
and the statement follows.

ii) From l'Hopitals rule   we get
\begin{alignat*}{4}
\lim_{\zeta\downarrow 0} \frac{\log(\frac{1}{G}\sum_j  e^{\zeta x_j} )}{\zeta}
=
\lim_{\zeta\downarrow 0} \Big(\frac{1}{G}\sum_j  x_je^{\zeta x_j} \Big)\Big(\frac{1}{G}\sum_j  e^{\zeta x_j}\Big)^{-1}
 = \frac{1}{G}\sum_j  x_j,
\end{alignat*}
implying
\begin{alignat*}{4}
  \lse(x)=    \frac {\log(G )}{\zeta}+ \frac {\log(\frac{1}{G}\sum_j  e^{\zeta x_j} )}{\zeta}=   \frac {\log(G )}{\zeta}+\frac{1}{G}\sum_j  x_j + o(1),
\end{alignat*}
for  $\zeta\downarrow0$.
\end{proof}

\begin{proof}[Proof of Proposition \ref{prop:zero}]

First  note that for  any $\beta\in \RR^p$ and  any $g\in \{1,\ldots,G\}$
  \begin{alignat}{4}\label{eq22}
 -\hat V_g(\beta)
  &= \beta^\top \hat \Sigma_g\beta-2\beta^\top \hat \Sigma_g b_g-\frac{2\beta^\top \bs X_g^\top \bs\epsilon_g}{n_g}\nonumber\\
  &=
  -V_{b_g}(\beta)
  +\beta^\top (\hat \Sigma_g-\Sigma)\beta -2\beta^\top (\hat \Sigma_g-\Sigma) b_g-\frac{2\beta^\top\bs X_g^\top \bs\epsilon_g}{n_g}\nonumber\\
   &\geq    -V_{b_g}(\beta)
   -\Vert \beta\Vert_1^2 D    -2\Vert\beta\Vert_1 \max_g\Vert b_g \Vert_1D -2\Vert\beta\Vert_1\delta
\end{alignat}
and correspondingly
\begin{alignat}{4}\label{eq23}
   -\hat V_g(\beta)&\leq   -V_{b_g}(\beta)
   +\Vert \beta\Vert_1^2 D    +2\Vert\beta\Vert_1\max_g \Vert b_g \Vert_1D +2\Vert\beta\Vert_1\delta.
\end{alignat}
 By assumption $\max_g\Vert b_g\Vert_1\leq\kappa$, and if also $\Vert\beta\Vert_1\leq\kappa$ we can write
\begin{alignat}{4}
\lse_\zeta(-V(\beta))-3\kappa^2 D -2\kappa \delta  &\leq\lse_\zeta( -\hat V(\beta))\label{eq24}\\
&\leq \lse_\zeta(-V(\beta))+3\kappa^2 D +2\kappa \delta,\label{eq25}
\end{alignat}
by \eqref{eq22} and \eqref{eq23} and the properties of $\lse_\zeta$.

Let $H$ denote the convex hull of $\{b_1,\ldots,b_g\}$.  By  Theorem 1 in \cite{meinshausen2015} we know that $b^\ast\in H$ and since  $\max_g\Vert b_g\Vert_1\leq\kappa$ we also have $\Vert b^\ast\Vert_1\leq \kappa$.
By \eqref{eq21}  it follows  that
 \begin{alignat}{4}\label{eq26}
    \lse_\zeta(-\hat V(\hat \beta_{smm}^\kappa))\leq  \lse_\zeta(-\hat V(b^\ast)).
\end{alignat}
Using \eqref{eq24} and  \eqref{eq25}  on respectively the left  and right hand side of \eqref{eq26},  yields
 \begin{alignat}{4}\label{eq27}
  \lse_\zeta(-V(\hat \beta_{smm}^\kappa))  &\leq
\lse_\zeta( - V(b^\ast))+6\kappa^2 D +4\kappa \delta.
\end{alignat}
Applying Lemma \ref{lemma:0} on both left  and right hand side of \eqref{eq27} finally yields
 \begin{alignat}{4}\label{eq28}
\max_g\{-V_{b_g}(\hat \beta_{smm}^\kappa)\}  &\leq
\max_g\{ - V_{b_g}(b^\ast)\}+6\kappa^2 D +4\kappa \delta+\frac{\log(G)}{\zeta}.
\end{alignat}

For the last statement,  note that for fixed $\beta\in \RR^p$,  since $b\mapsto-V_b(\beta)$ is affine
\begin{alignat}{4}\label{eq29}
  \max_{g} \{-V_{b_g}(\beta)\}=\max_{b\in H} \{-V_{b}(\beta)\}.
\end{alignat}
 An implication of Theorem 1 in \cite{meinshausen2015}  is
  $(b^\ast)^\top\Sigma b\geq (b^\ast)^\top\Sigma b^\ast, \forall b\in H$, which combined with
   \eqref{eq29} shows
\begin{alignat}{4}\label{eq30}
\max_{g}\{-V_{b_g}(b^\ast)\}
 = \max_{b\in H}\{(b^\ast)^\top \Sigma b^\ast- 2(b^\ast)^\top \Sigma b\}
 =-(b^\ast)^\top \Sigma b^\ast.
\end{alignat}
 Using  $b^\ast\in H$ and \eqref{eq29}, on the left hand side of \eqref{eq28},   and using \eqref{eq30} on the right hand side of \eqref{eq28}, gives us
 \begin{alignat*}{4}
 (\hat \beta_{smm}^\kappa)^\top \Sigma \hat \beta_{smm}^\kappa- 2(\hat \beta_{smm}^\kappa)^\top \Sigma b^\ast\leq-(b^\ast)^\top \Sigma b^\ast+6\kappa^2 D +4\kappa \delta+\frac{\log(G)}{\zeta},
\end{alignat*}
which can be rearranged to yield the statement.
\end{proof}

To prove Proposition \ref{prop:one} we  need the following technical lemma.
\begin{thm_lemma}\label{lemma:one}
 Assume  $\sum_i w_i=1$ and  $h_i\in \RR^p	$, $i \in \{1,\ldots,G\}$, $G\in \NN$. Then
 \begin{alignat*}{4}
     \sum_{i} w_ih_i \Big(h_i^\top -  \sum_{j} w_jh_j^\top  \Big)=\sum_i \sum_{j>i}w_iw_j (h_i - h_j) (h_i - h_j)^\top.
\end{alignat*}
\end{thm_lemma}
 \begin{proof}
 First note that since $1 - w_i = \sum_{j \neq i} w_j$
 \begin{alignat*}{4}
   \sum_{i} w_ih_i\Big(h_i^\top -  \sum_{j} w_jh_j^\top  \Big)&=
   \sum_{i} w_ih_i\Big((1-w_i) h_i^\top -  \sum_{j\neq i}w_jh_j^\top  \Big)\\
   &=  \sum_{i} \sum_{j\neq i} w_iw_j  h_i(h_i - h_j)^\top.
   \end{alignat*}
Letting $a_{i,j} = w_iw_j h_i (h_i - h_j)^\top$ we find that
\begin{alignat*}{4}
  \sum_{i} \sum_{j\neq i} a_{i,j}
    &= \sum_{i}\sum_{j>i} a_{i,j} +
   \sum_{i}\sum_{i > j } a_{i,j} \\
&= \sum_{i}\sum_{j>i} a_{i,j} +
   \sum_{j}\sum_{i > j} a_{i,j} \quad (\text{interchange summation order})\\
&= \sum_{i}\sum_{j>i} a_{i,j} +
   \sum_{i}\sum_{j>i} a_{j,i}  \quad (\text{relabel summation indices})\\
  &=\sum_i \sum_{j>i} w_iw_j (h_i - h_j) (h_i - h_j)^\top,
\end{alignat*}
where we used that $a_{j,i} = - w_iw_j h_j (h_i - h_j)^\top$.
\end{proof}
\begin{proof}[Proof of Proposition \ref{prop:one}]
First  it is straightforward to compute the gradient of the loss
\begin{alignat}{4}
\nabla l_\zeta(\beta)&   =\frac{\sum_{j=1}^Ge^{\zeta h_j(\beta)}  \nabla  \zeta h_j(\beta)}{\zeta\sum_{j=1}^G e^{\zeta h_j(\beta)} }\nonumber \\
&   =e^{- \log(\sum_{j=1}^G e^{\zeta h_j(\beta)})}\sum_{j=1}^Ge^{\zeta h_j(\beta)}  \nabla   h_j(\beta)\nonumber\\
&= \sum_{j=1}^Gw_{j,\zeta}(\beta) \nabla  h_j(\beta).\label{eq17new}
\end{alignat}
Then since
\begin{alignat*}{4}
w_{j,\zeta}(\beta) =\frac{e^{\zeta h_j(\beta)} }{\sum_{i=1}^G e^{\zeta
    h_i(\beta)} } \geq 0
   \end{alignat*}
we see that $\sum_{j} w_{j,\zeta}(\beta) = 1$ and conclude that the weights $w_\zeta(\beta)$ are convex for any $\beta$ and $\zeta$.

Differentiating \eqref{eq17new}  gives us
\begin{alignat*}{4}
\nabla^2 l_\zeta(\beta)
&= \sum_{j=1}^G  \nabla  h_j(\beta)\nabla w_{j,\zeta}(\beta)^\top
+ \sum_{j=1}^G w_{j,\zeta}(\beta) \nabla^2  h_j(\beta).
\end{alignat*}
Using the definition of $w_{j,\zeta}$ and \eqref{eq17new} the first term is equal to
\begin{alignat*}{4}
\sum_{j=1}^G w_{j,\zeta}(\beta) \nabla & h_j(\beta)\Big(\nabla
h_j(\beta)^\top- \sum_{i=1}^G w_{i,\zeta}(\beta)\nabla  h_{i}(\beta)^\top\Big)\\
&=\sum_{j=1}^G \sum_{i>j} w_{j,\zeta}(\beta)w_{i,\zeta}(\beta) (\nabla  h_j(\beta)-\nabla  h_{i}(\beta))(\nabla  h_j(\beta)-\nabla  h_{i}(\beta))^\top
\end{alignat*}
where the equality follows from  Lemma \ref{lemma:one} since $(w_{j,\zeta}(\beta))_j$ are convex weights.

For a twice continuously differentiable function $f$ it holds that $f$   is strongly convex with parameter  $\nu>0 $ if and only if  $	\nabla^2f -\nu I$ is positive semi definite. Assuming all $h_i$ are convex and at least one is $\nu$-strongly convex it follows directly from \eqref{eq10new} that $l_\zeta$ is also  $\nu$-strongly convex.

 Finally, the Hessian of  $e^{  l_\zeta(\beta)}$  is
\begin{alignat*}{4}
\nabla^2e^{  l_\zeta(\beta)}  = \nabla^2 l_\zeta(\beta)e^{  l_\zeta(\beta)}+\nabla_\beta  l_\zeta(\beta)\nabla_\beta  l_\zeta(\beta)^\top e^{  l_\zeta(\beta)}
\end{alignat*}
where $\nabla_\beta  l_\zeta(\beta)\nabla_\beta  l_\zeta(\beta)^\top e^{  l_\zeta(\beta)}$ is positive semi-definite for all $\beta$. Letting    $m=\min_\beta l_\zeta(\beta)\in \RR $, we have  $ e^{ l_\zeta(\beta)}\geq e^{ m}>0$ for all $\beta$ and must have   $\nabla^2e^{  l_\zeta(\beta)} -\tilde \nu I$ is positive semi definite for some $\tilde \nu >0$, showing that  $e^{ l_\zeta}$ is strongly convex.
\end{proof}

\begin{proof}[Proof of Corollary \ref{coro:one}]
Let  $\Vert\cdot\Vert_d $ denote  the 2-norm on $\RR^{d}$  and   $A$  be a $d_1\times d_2$ matrix. Then
$ \vvvert A \vvvert_{d_1,d_2}:=\sup_{v:\Vert v\Vert_{d_2}=1}\Vert Av\Vert_{d_1}  $
is the sub-multiplicative matrix (operator) norm induced by the 2-norms on $\RR^{d_1}$ and $\RR^{d_2}$.   For $a\in \RR^{d_1}$ and $b\in \RR^{d_2}$  note that  $\vvvert a \vvvert_{d_1,1}=\Vert a\Vert_{d_1}$  (Cauchy-Schwarz)  and we get
\begin{alignat*}{4}
  \vvvert ab^\top\vvvert=\sup_{v:\Vert v\Vert_{d_2}=1}\Vert ab^\top v\Vert_{d_1}
  =\sup_{v:\Vert v\Vert_{d_2}=1}\Vert a\Vert_{d_1}  \vert b^\top v\vert
  =\Vert a\Vert_{d_1}\Vert b\Vert_{d_2}.
\end{alignat*}

Now suppressing subscripts  observe that $\vvvert \nabla ^2h_g(\beta) \vvvert= 2 \vvvert \bs X^\top \bs X\vvvert /m$  and  $\nabla h_i(\beta) - \nabla h_j(\beta) =2\bs X^\top (\bs Y_i-\bs Y_j)/m$.
 Then by Proposition \ref{prop:one} it follows that
\begin{alignat*}{4}
\vvvert \nabla^2 l_\zeta(\beta)\vvvert  &\leq \sum_{i}\sum_{j>i}  w_{i,\zeta}(\beta)w_{j,\zeta}(\beta) \vvvert (\nabla  h_i(\beta)-\nabla  h_{j}(\beta))(\nabla  h_i(\beta)-\nabla  h_{j}(\beta))^\top\vvvert\nonumber\\
	&\phantom{\leq}+ \sum_{j} w_{j,\zeta}(\beta) \vvvert\nabla^2  h_j(\beta)\vvvert
\nonumber \\
&= \frac{4}{m^2}
\sum_{i}\sum_{j>i}
w_{i,\zeta}(\beta)w_{j,\zeta}(\beta) \Vert  \bs X^\top (\bs Y_i-\bs Y_j) \Vert^2+ \frac{2 \vvvert \bs X^\top \bs X\vvvert }{m}
\nonumber\\
&\leq  \frac{4}{m^2}\max_{i,j}\Vert \bs X^\top (\bs Y_i-\bs Y_j) \Vert^2+ \frac{2 \vvvert \bs X^\top \bs X\vvvert }{m}
\nonumber\\
&\leq   \frac{4 \vvvert \bs X^\top \bs X\vvvert }{m^2}\Big(\max_{i,j}  \Vert \bs Y_i-\bs Y_j\Vert^2  +  \frac{m}{2}\Big)
\end{alignat*}
 using the properties of the matrix norm.  By the mean value theorem it follows that $\nabla l_\zeta$  is  Lipschitz continuous with the claimed bound.
\end{proof}

\begin{proof}[Proof of Proposition \ref{prop:two}]
 If we can show that   Assumption A.1 from \cite{chen2016}  holds for
 the soft maximin problem \eqref{eq13} we can use Theorem A.1 in
 \cite{chen2016} (or Lemma 4 in \cite{wright2009}) to show that
 the sequence has an accumulation point.  Theorem 1 in \cite{wright2009} then establishes this accumulation point as a critical point for $F_{\zeta}=l_\zeta+\lambda J$.

  Let $\Delta>0$,    $\beta_0\in \RR^p$  , and define the set
  \begin{alignat*}{4}
  A_0&=\{\beta:F_{\zeta}(\beta)\leq F_{\zeta}(\beta_0)\} \\
  A_{0,\Delta}&=\{\beta:\Vert \beta-\beta'\Vert\leq \Delta, \beta'\in A_0\}.
\end{alignat*}

 A.1(i): $ l_\zeta$ is $\nu$-strongly convex by Proposition \ref{prop:one}  and since $J$ is assumed convex it follows that $F_\zeta$ is strongly convex. So  $A_0$ is compact hence $A_{0,\Delta}$ is compact as a closed neighbourhood of $A_0$. As $ l_\zeta$  is $C^{\infty}$ everywhere,   $\nabla  l_\zeta $ is Lipschitz on $A_{0,\Delta}$.

 A.1(ii): Is satisfied by assumptions on $J$.

 A.1(iii): Clearly $F_{\zeta}\geq0$. Furthermore $F_{\zeta}$ is continuous hence uniformly continuous on the compact set $A_0$.

A.1(iv)  $\sup_{\beta\in A_0}\Vert\nabla l_\zeta \Vert<\infty$ as $A_0$ is compact and $\nabla l_\zeta$ is continuous.
 Moreover, $\sup_{\beta\in A_0}\Vert J\Vert<\infty$ as $A_0$ is compact and $J $ is continuous.  Finally, also $\inf J=0$.
\end{proof}

\section{Brain imaging data}\label{sec:neuron}
The neuronal activity recordings were obtained using
 voltage-sensitive dye imaging (VSDI) in an experiment previously described in
 \cite{roland2006}.  In short  part of the visual cortex of a live ferret was
 exposed and stained with voltage-sensitive dye. Changes in membrane potential  affects the dye and alters its fluorescence. The neuronal activity
 is recorded indirectly in terms of changes in the  fluorescence using 464 photodiode channels  organized in a two-dimensional (hexagonal) array.  By padding with zeros  the 464 channels were  mapped to a $25\times25$ matrix. We note the padding is chosen as the data is centred around zero implying the analysis is not altered by this manipulation.  Alternatively observation weights can be used at a computational cost.
 During the trial (625 ms) an image was recorded every $0.6136$ ms.  For 250 ms of the trial a visual stimulus, a white square on a grey screen,  was presented to  the  ferret.   A total of $G=275$  trials  were recorded across 13 different ferrets.

Several sources of heterogeneity are potentially present in the raw data:
\begin{enumerate}
\item\label{list:iv}   The heart beat
affects the light emission by expanding the blood vessels in the brain,
creating  a cyclic heart rate dependent artefact. A changing heart rate over
trials for   one animal (fatigue) as well as differences in heart rate  between animals
will cause heterogeneity in the data.
\item\label{list:ii}  Spatial inhomogeneities can  arise due to  differences in the cytoarchitectural borders between the animals causing misalignment problems.
\item\label{list:iii}   The VSDI technique is very sensitive, see  \cite{grinwald2002}.   Even small changes in the experimental surroundings could affect the recordings and create heterogeneity.
\item\label{list:v} Differences between animals in how they
  respond to the visual stimulus.
\end{enumerate}

A trial with no  visual stimulus (baseline), was recorded right before recording the stimulus trial.  By   aligning the baseline and stimulus trial, using an electrocardiography recording, the two recordings were subtracted to  remove the heart rate artefact. We  use this preprocessed data in the experiment.

\begin{sidewaysfigure}
\centering
\includegraphics[scale=0.58]{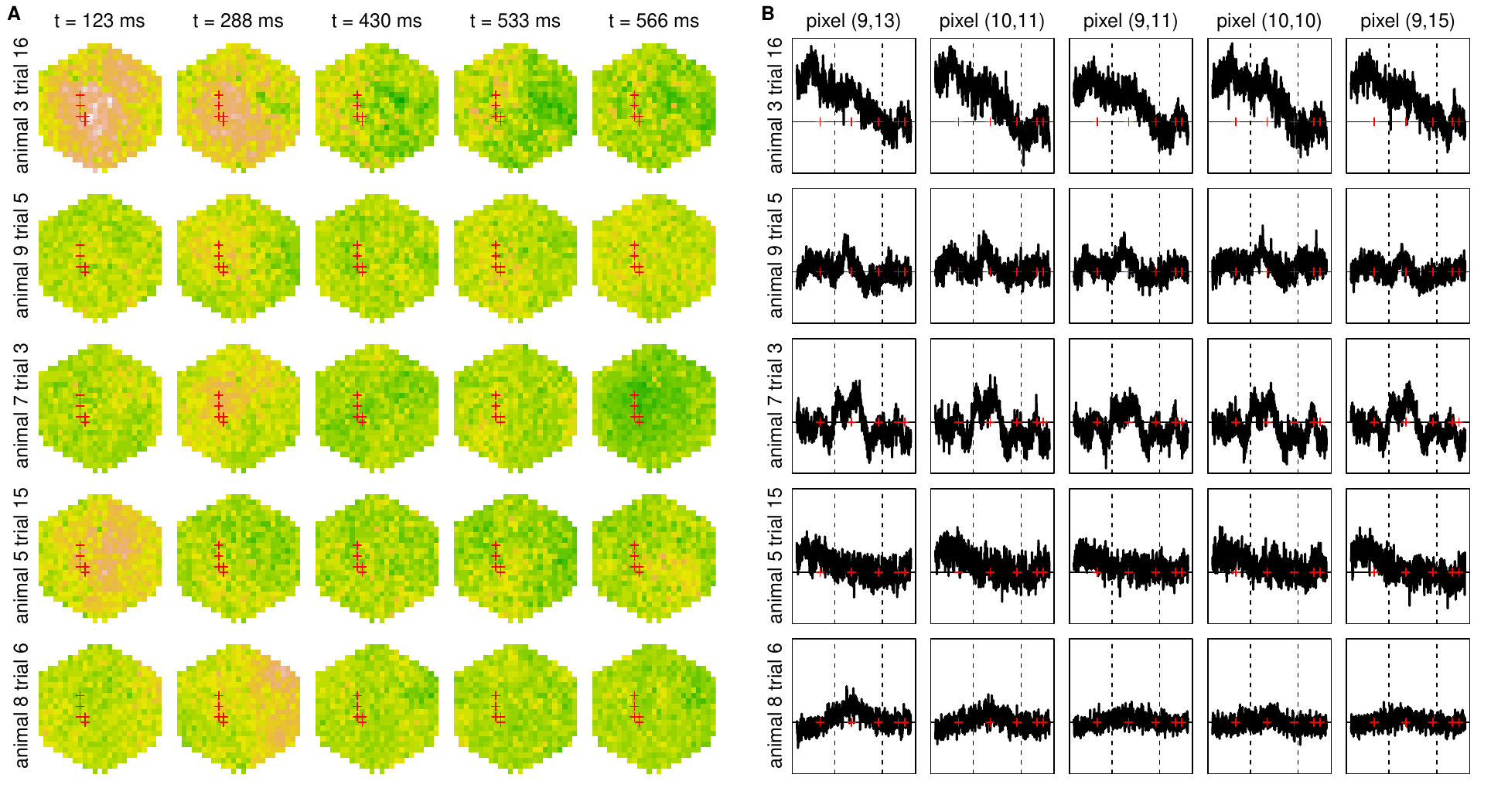}
\caption{A: Spatial plots for five  trials and five  time points.  The red crosses indicate the pixel time series shown in B.
      B: Temporal  plots for  the five same trials and five pixels indicated with red crosses in part A.  The dashed lines indicate stimulus start and stop and  red crosses indicate the time points plotted in  part A. }
           \label{fig:5}
    \end{sidewaysfigure}

Figure \ref{fig:5}  shows
recordings for five trials in  the temporal dimension (panel A) and spatial dimension (panel B).
 Note that following the
onset of the visual stimulus after 200 ms (first dashed line), the recordings are expected to
show the result of a depolarization of the neuronal cells.
Visual inspection of Figure \ref{fig:5} however  does not reveal a clear stimulus response in  every trial. We note   the presence of   variation  that seems to be specific to the trial and  could reflect the heterogeneity   listed above.

\subsection{Experiment setup}
We use the array-tensor model from  Section \ref{subsec:atsmooth} with  $p_1=p_2=9$ B-spline functions  in each  spatial dimension  and $p_3=80$
B-splines in the temporal dimension. This gives us  a
model with marginal design matrices $\Phi_1$, $\Phi_2$, and $\Phi_3$, of sizes $25\times 9$, $25\times 9$ and $977\times 80$ respectively, given by the
B-splines evaluations over the marginal domains. The  model
has  $p =6480$  parameters.

We let one fold consist of all data from 2 out of the 13 animals.  The model is trained  on the fold and tested on   data from the remaining 11 animals, for each method and each value of $\lambda$.   We repeat this procedure $N:=\binom{13}{2}=78$ times, yielding 78 fitted models and corresponding test metrics for each method and each value of $\lambda$.
Since the number of trials is not constant across animals the number of groups in each fold ranges from 23 to 80 (average is 42) giving us $14,044,375 $ to $48,850,000$ (average $25,646,250$) observations in each fold.

 \subsection{Experimental results}\label{exp_res_real}
\begin{figure}[H]
\begin{center}
\includegraphics[scale = 0.35]{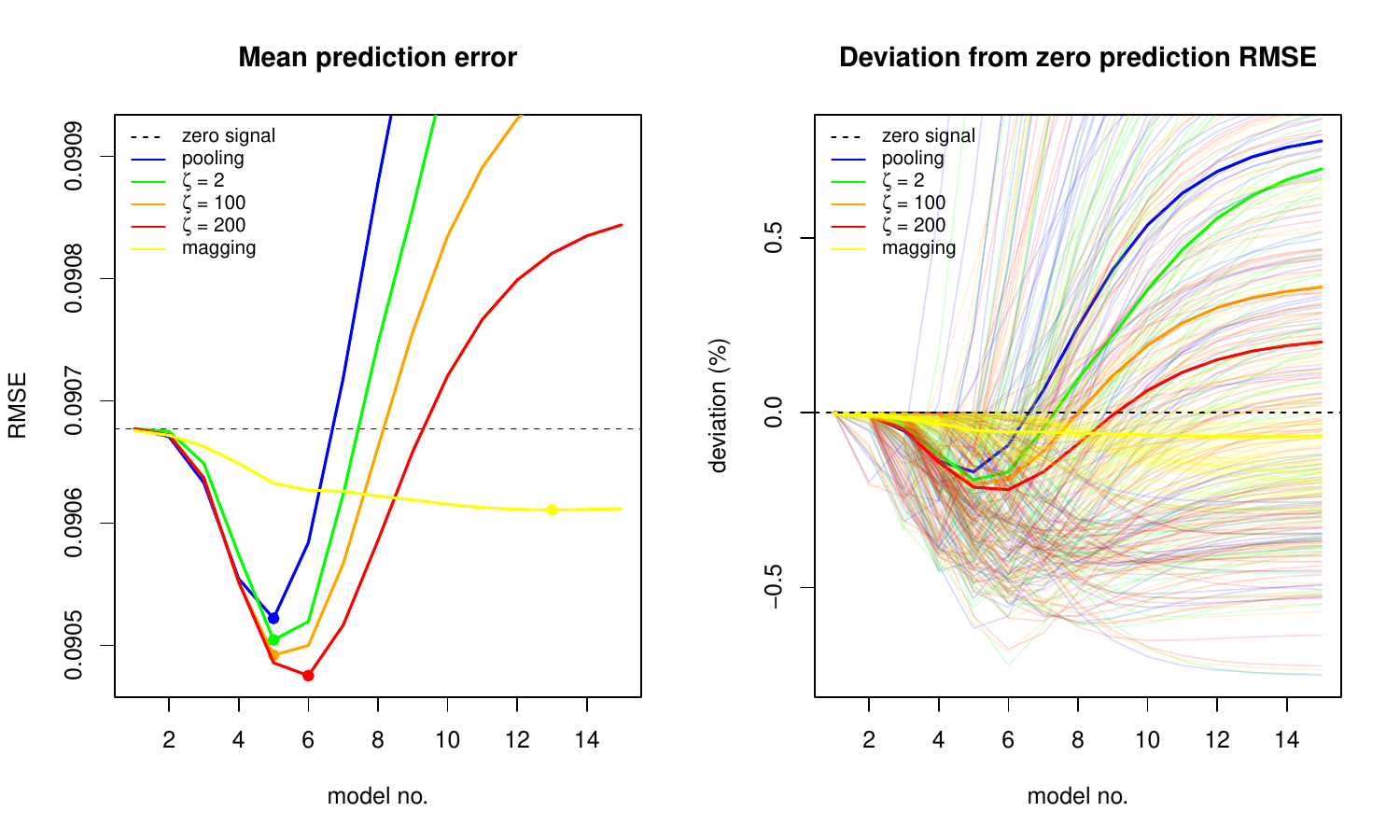}
\includegraphics[scale = 0.3]{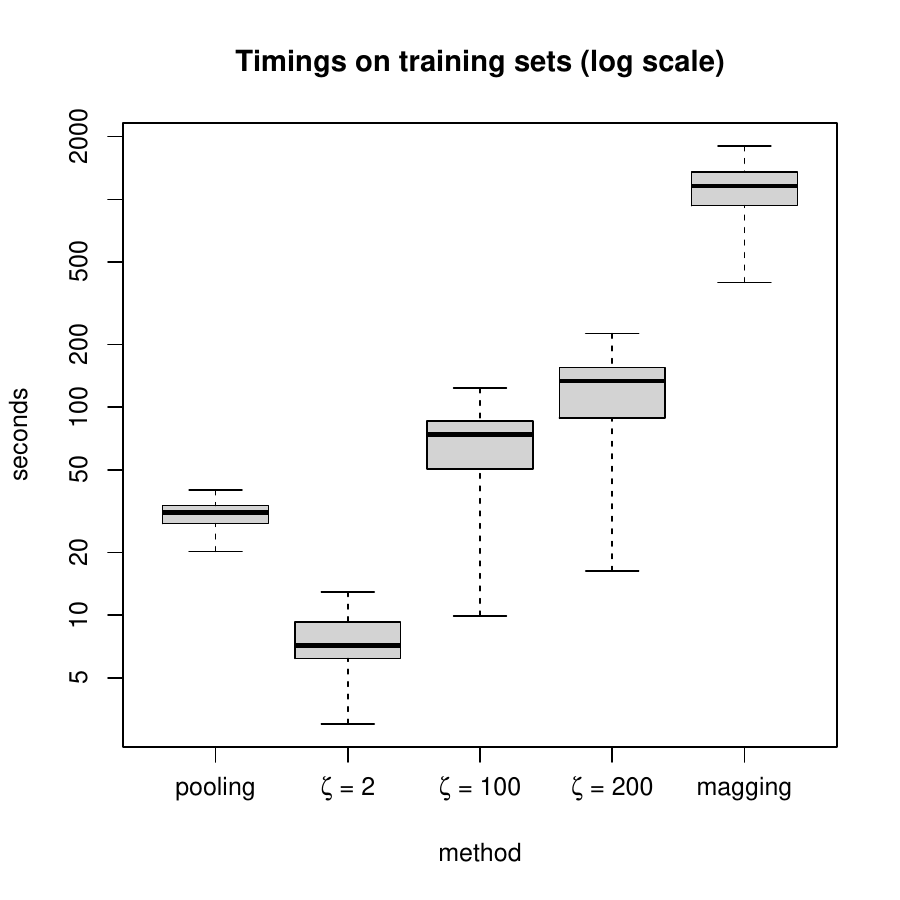}
\caption{Top left: Average RMSE across  78  test sets as a function of model complexity (model no.).
 Zero signal (dashed), pooling  (blue), soft maximin   $\zeta\in \{2,100,200\}$ (resp. green, orange and red ) and magging (yellow). 
 Top right:  Deviation  in RMSE relative to  the zero prediction  for each method and on each test set (thin lines), as a function of model complexity.  Averages are indicated with a thick line of the same color. Bottom:  Run times (log scale) for the 78 training sets for each method.}
\label{fig:6}
\end{center}
\end{figure}
\newpage
 From the left display in Figure \ref{fig:6} we see that on average soft maximin with $\zeta =200$ (model no. 6) achieves the lowest over all out of sample RMSE. The low $\zeta$ estimators, pooled (model no. 5)  and $\zeta=2$ (model no. 5) perform somewhat worse on average but still achieves significant reduction compared to the zero prediction. The approximate maximin estimator, magging estimator (model no. 13), performs the worst  on this data in terms of RMSE.

Looking at Figure\ref{fig:6}  the low $\zeta$ methods show  more variability than the high $\zeta$ methods and in particular are more prone to make predictions that are worse than the zero prediction. However the picture  is not as clear as on the simulated data.  We note that the  magging estimator is quite consistently better than the zero prediction but not much,  possibly reflecting the conservative nature of the hard maximin method.

Figure \ref{fig:6} also summarizes the timings   for each method. Notably all the soft maximin estimators  (8.9 s, 15.8 s and 18.7 s) outperform the pooled estimator (26.4 s)   while also yielding better prediction accuracy (Figure \ref{fig:6}). The magging estimator  (931.4 s) suffers from having to compute individual fits for each group in the training set, making the the method orders of magnitudes slower in this case, without obtaining better accuracy. Note that  the task alone of maximin aggregating the individual group estimates, by solving the associated quadratic programming problem, took on average 45 s. So even if fully parallelized the magging estimator is still computationally more demanding than the softmaximin estimator on this data set.

\section{ Washington DC bike data}\label{sec:bike_app}

 Here we   show how to systematically determine the soft maximin parameter $\zeta$. 
 
  Owing to the temporal dependence in the data we will use a rolling cross validation scheme  to systematically tune $\zeta$.  We  do this by training the model \eqref{eq20new} on each set of six consecutive months  and testing  on the  six following months.  Following the experiment in section \ref{sec:bike} we perform this rolling window CV in two ways; i)  rolling forward from Jan 2011 to Dec  2012 and ii) rolling   backwards from Dec 2012 to Jan 2011. 
  
  For i) and ii) respectively we  then have 13 training and test pairs. We compute  the soft maximin estimator on each of   these training set for 50  values of $\zeta$ that range exponentially between 0.0001 and 0.3.  On the corresponding test set we compute the mean squared prediction error. 
  
\begin{figure}[H]
\begin{center}
\includegraphics[scale = 0.4]{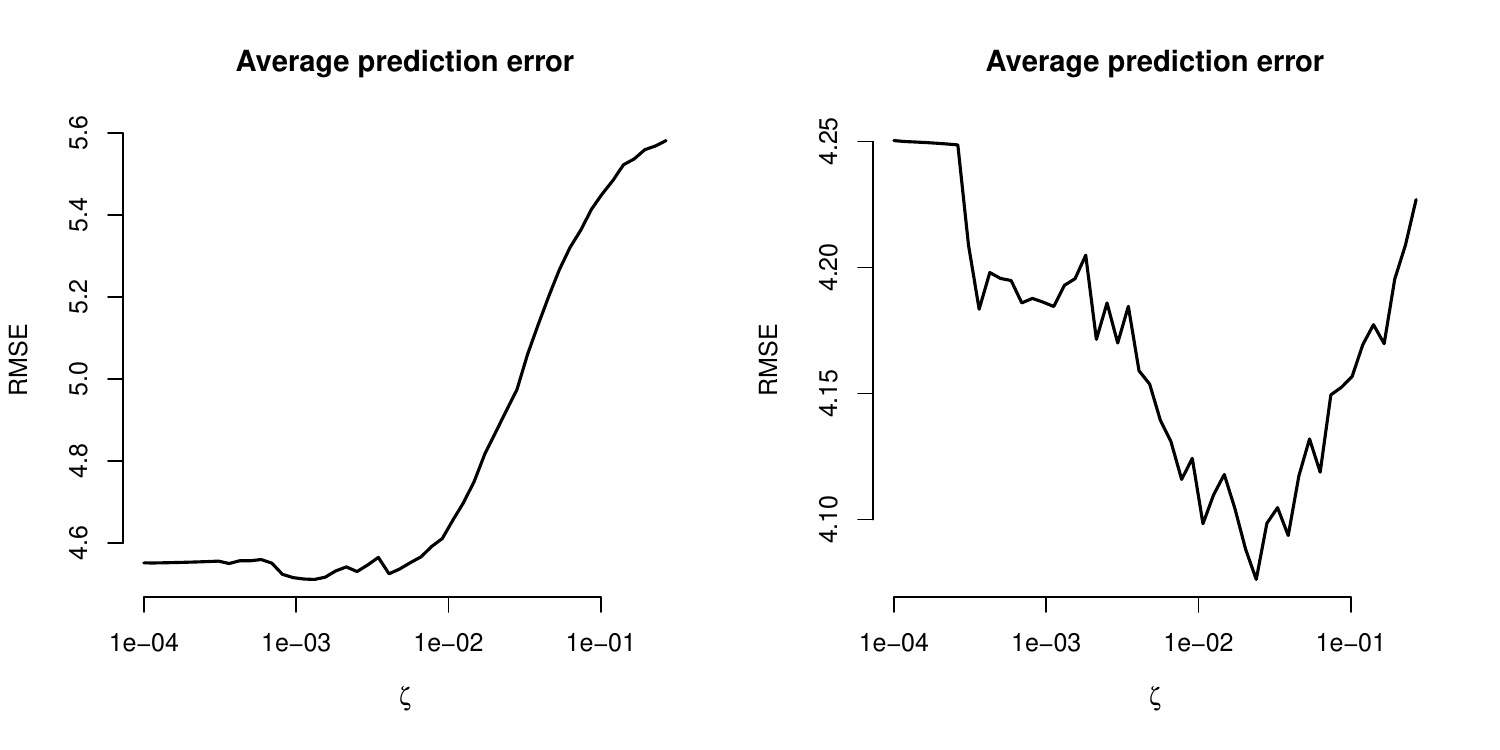}
\caption{Left: Average prediction error as a function of $\zeta$ obtained by the forward rolling cross validation procedure in experiment i). Right: Average prediction error as a function of $\zeta$ obtained by the backward rolling cross validation procedure in experiment ii)}
\label{fig:9}
\end{center}
\end{figure}

  Figure \ref{fig:9} shows the    average  prediction error (RMSE ) as a function of $\zeta$ for the forward rolling scheme i) and the backward rolling scheme ii) respectively. In line with section \ref{sec:bike} in i) we see that low $\zeta$ values i.e. pooled OLS  gives better predictions in terms of RMSE than higher values. For the experiment in ii) however $\zeta$ values around 0.03 yields the minimum prediction error. 

  We note that any conclusion will depend  on the nature of the heterogeneity in the data as well as on how the cross validation is carried out, i.e. how the model is trained and tested.  For the specific bike data set  heterogeneity is  not very pronounced and is easily explained  in terms of  increasing utilization of the bike sharing scheme. This causes an optimistic method like pooling to perform better over time than a conservative method like maximin. However we  observe that the hard maximin (high $\zeta$) seems suboptimal in both  experiments i) and ii) highlighting the benefit of computing a range of soft maximin  estimators for a given data set.

\subsection{Alternative models and grouping}

\begin{figure}[H]
\begin{center}
\includegraphics[scale = 0.4]{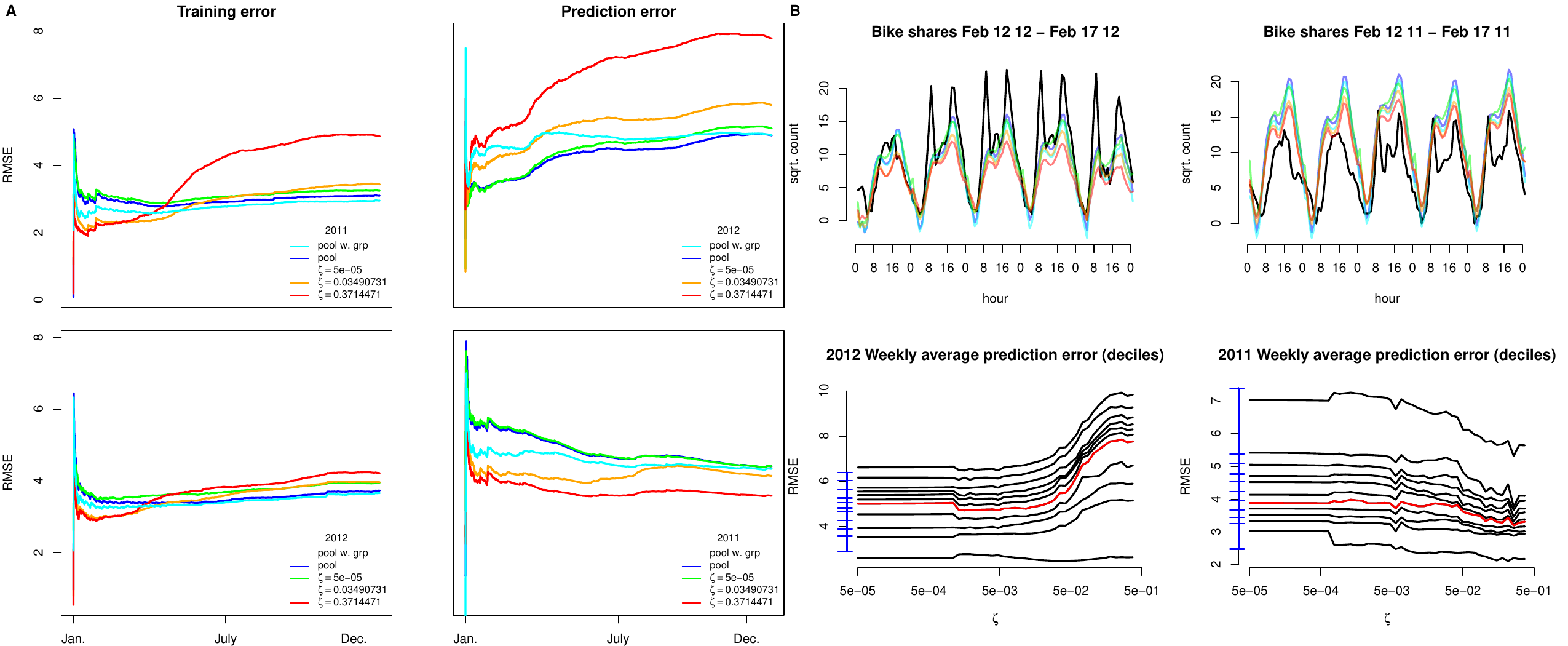}
\caption{Results from fitting a model that includes temperature and humidity as covariates and using  month as grouping.}
\label{fig:10}
\end{center}
\end{figure}

Figure \ref{fig:10} shows the results with  temperature  \verb+temp+ and humidity \verb+hum+ added to \eqref{eq20new}. Results for this model are similar to the those  in section \ref{sec:bike} though with less pronounced gain in prediction robustness (Figure \ref{fig:10} B Bottom right).

\begin{figure}[H]
\begin{center}
\includegraphics[scale = 0.4]{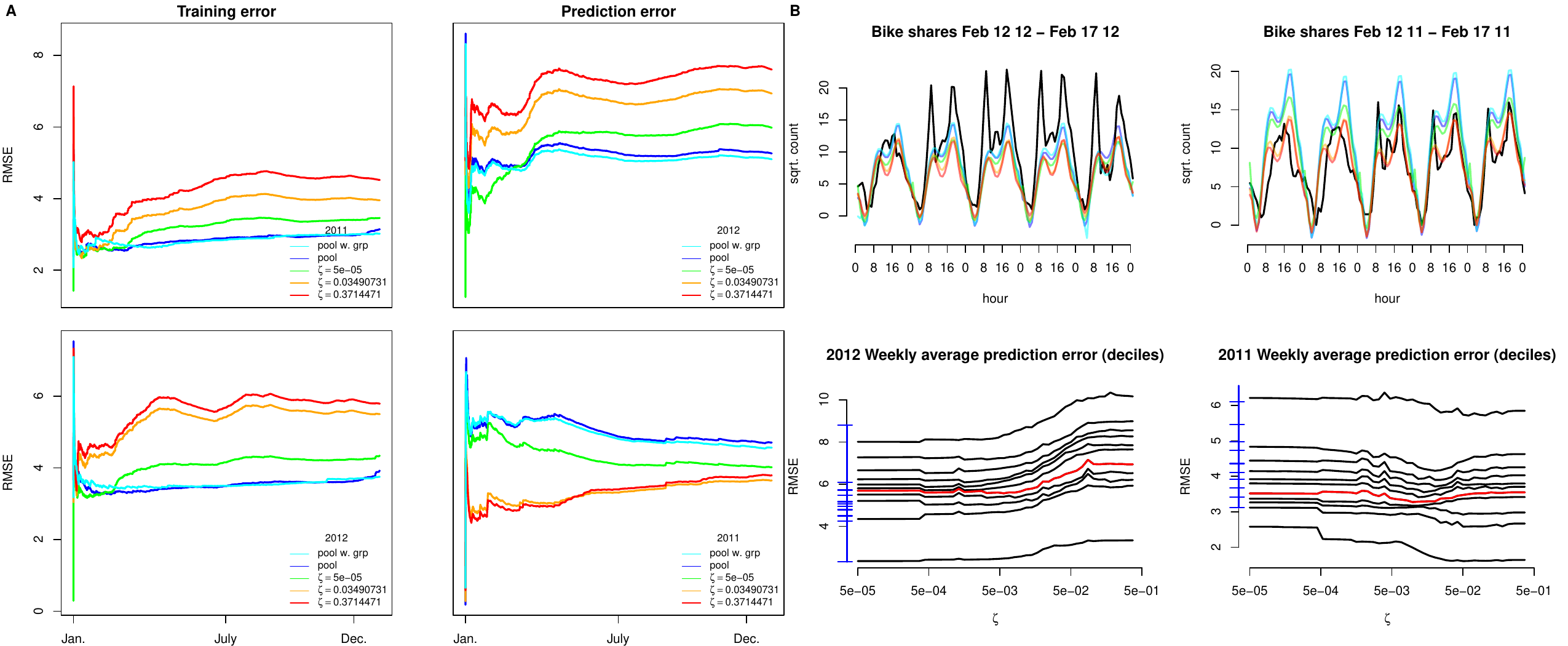}
\caption{Results from fitting a model that use \texttt{weathersit} as grouping.}
\label{fig:11}
\end{center}
\end{figure}

We also group the data according to  \verb+weathersit+ and fit the model
 \begin{alignat}{4}\label{eq34}
\sqrt{\texttt{cnt}_i}=\alpha_0+\sum_{j=1}^{10}\alpha_j \phi_j(\texttt{hr}_i)+\sum_{j=1}^{5}\beta_j \phi_j(\texttt{weekday}_i)+
 \sum_{j=1}^5 \gamma_j  \phi_j(\texttt{mnth}_i) +\epsilon_i.
\end{alignat}
With this setup we obtain slightly higher 2011 prediction accuracy but less  robust predictions (Figure \ref{fig:11} A and B bottom right). 

\begin{figure}[H]
\begin{center}
\includegraphics[scale = 0.4]{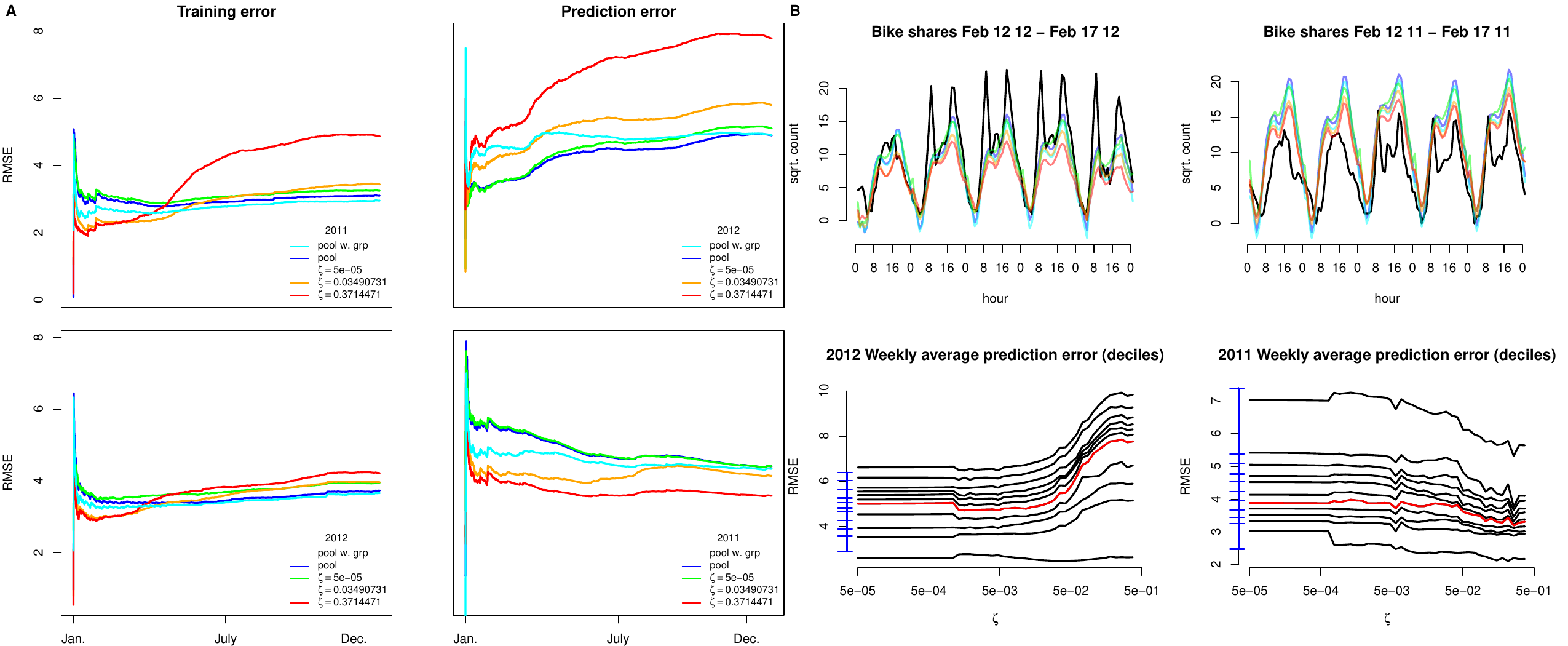}
\caption{Results from fitting a  model that includes temperature and humidity and use \texttt{weathersit} as grouping.}
\label{fig:12}
\end{center}
\end{figure}

Finally we also add \verb+temp+ and \verb+hum+ to \eqref{eq34}. For this extended model   the soft maximin estimator gives seemingly better 2011 predictions than  the pooled estimator, see Figure \ref{fig:12}.
   
 \newgeometry{left=1cm, right=1cm
 }

 \restoregeometry

\bibliography{bibliotek}

\end{document}